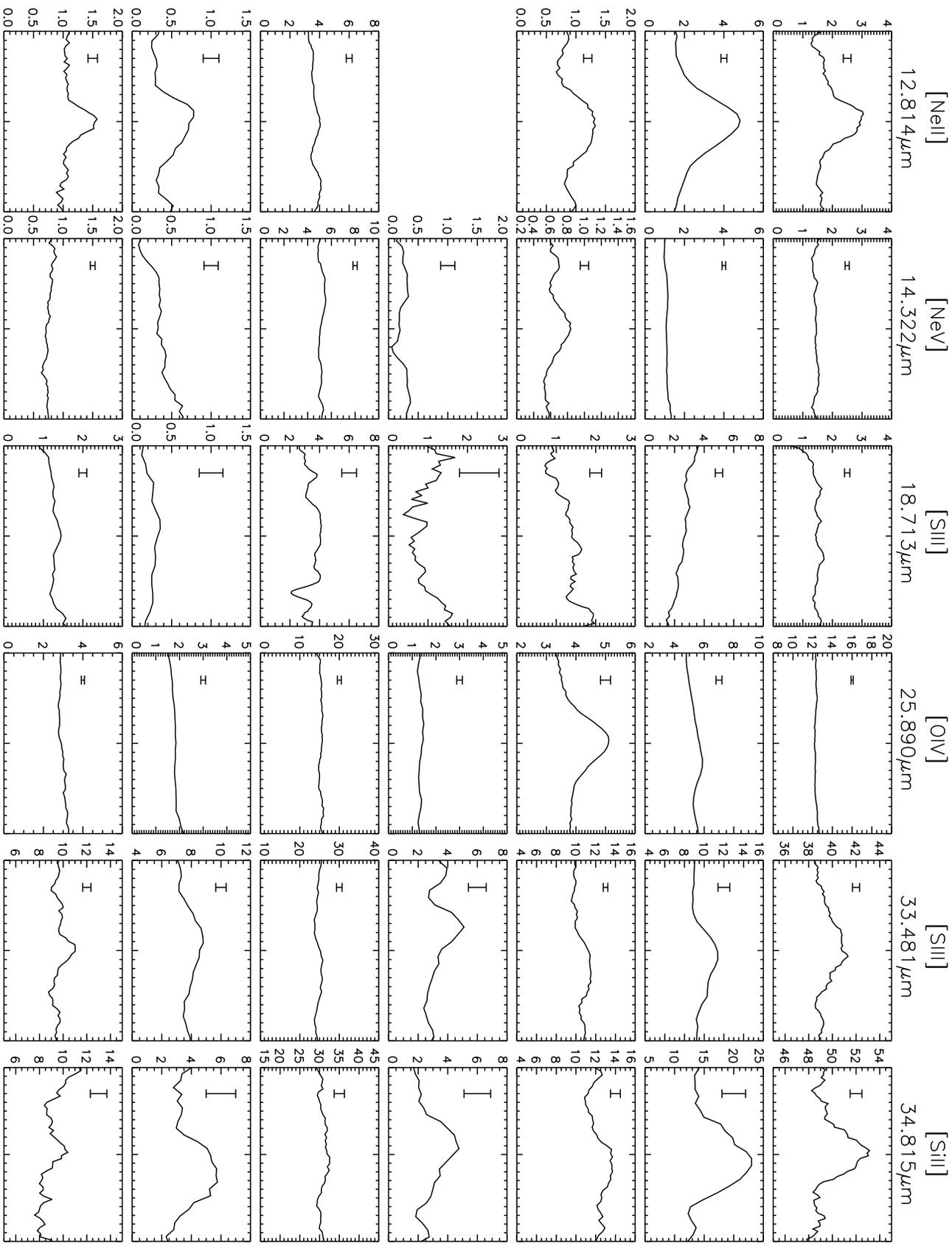

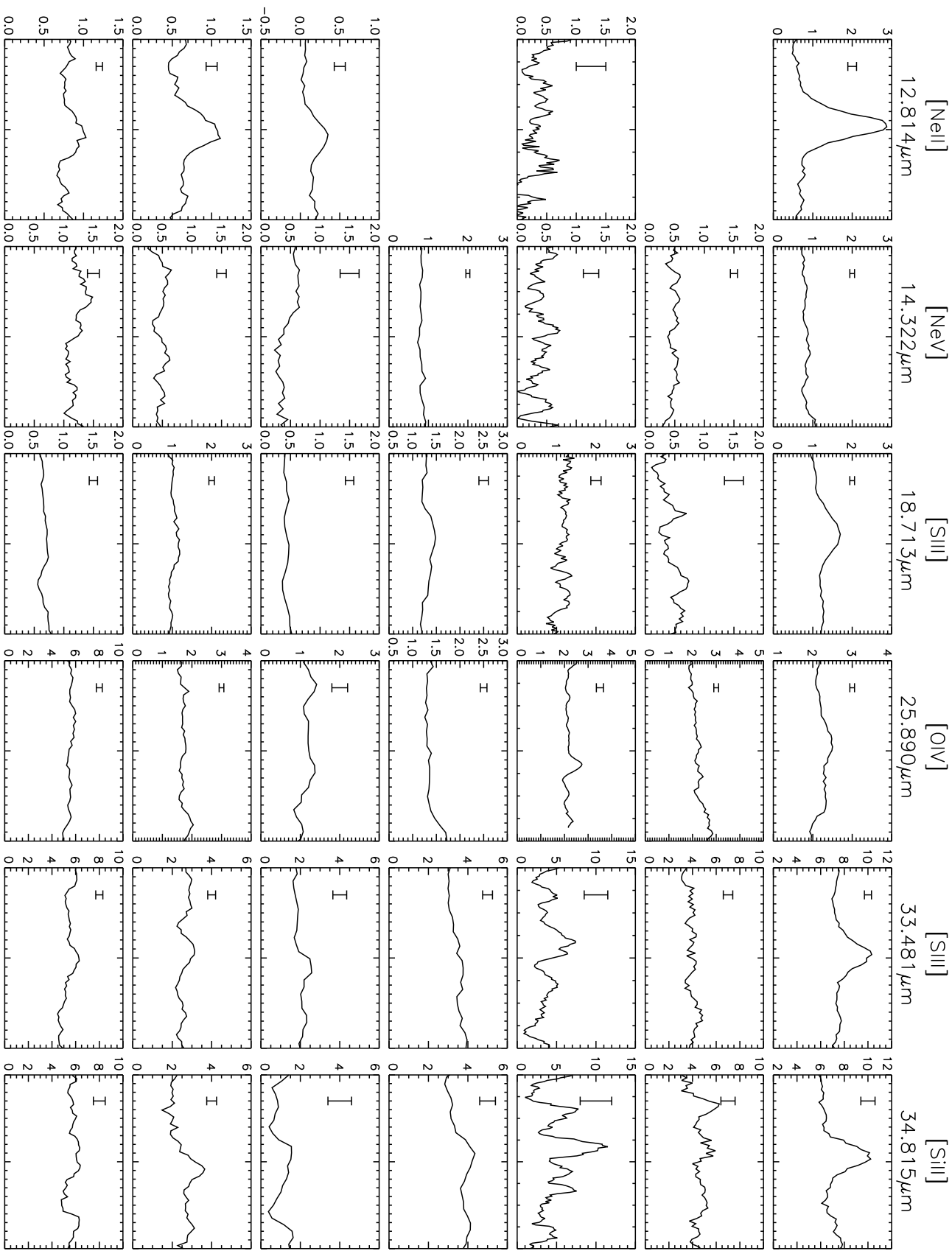

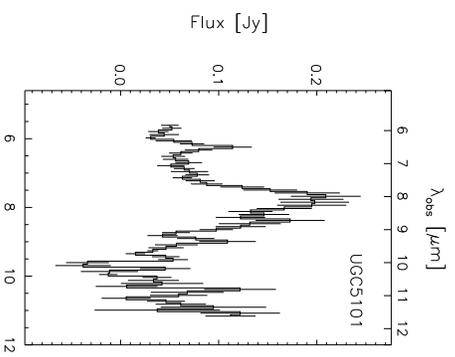

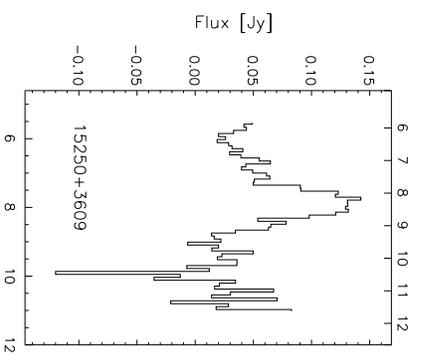

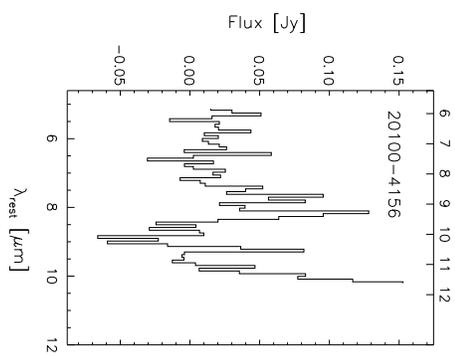

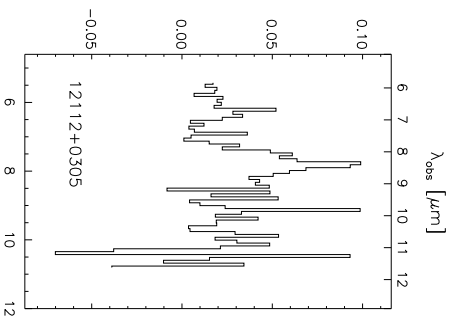

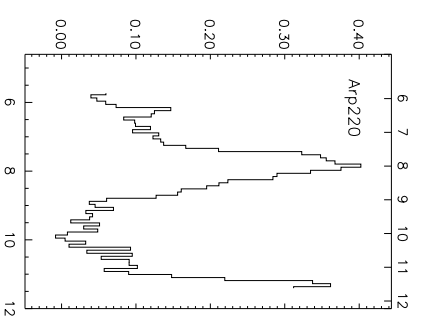

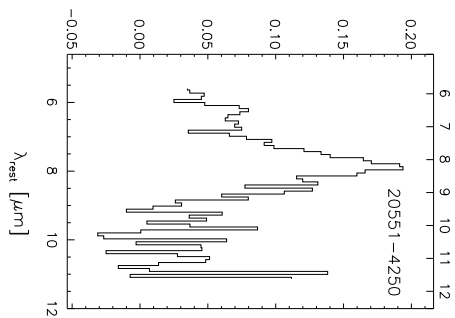

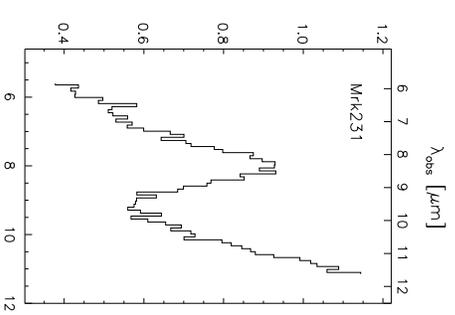

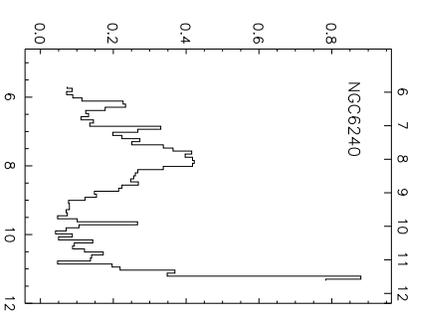

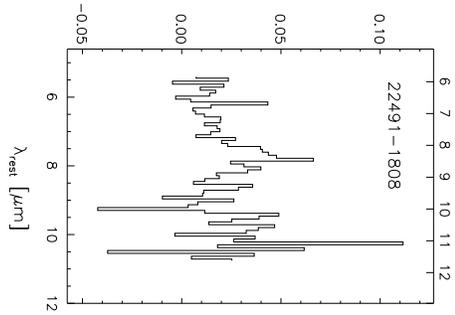

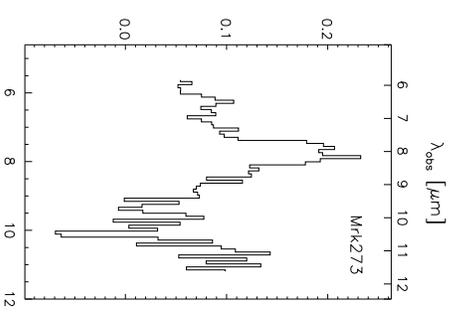

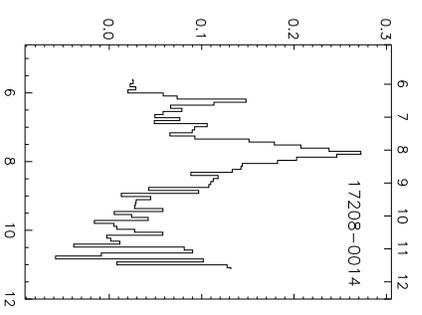

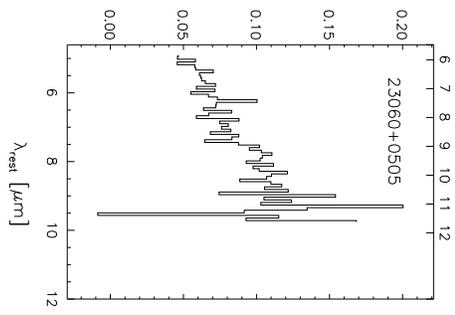

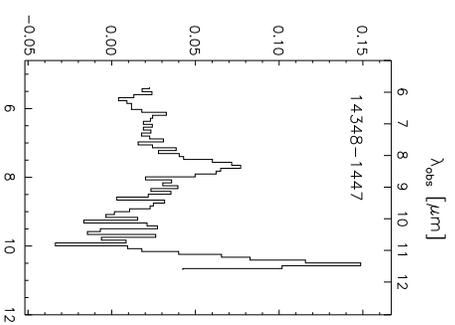

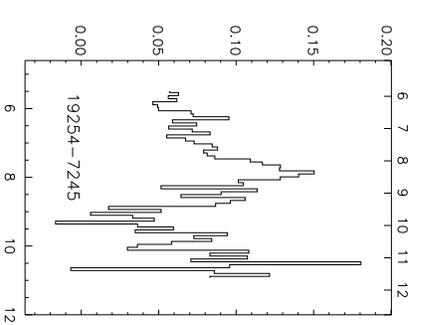

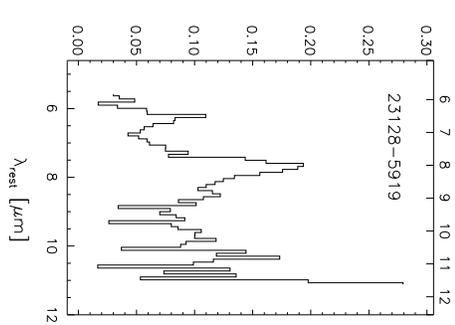

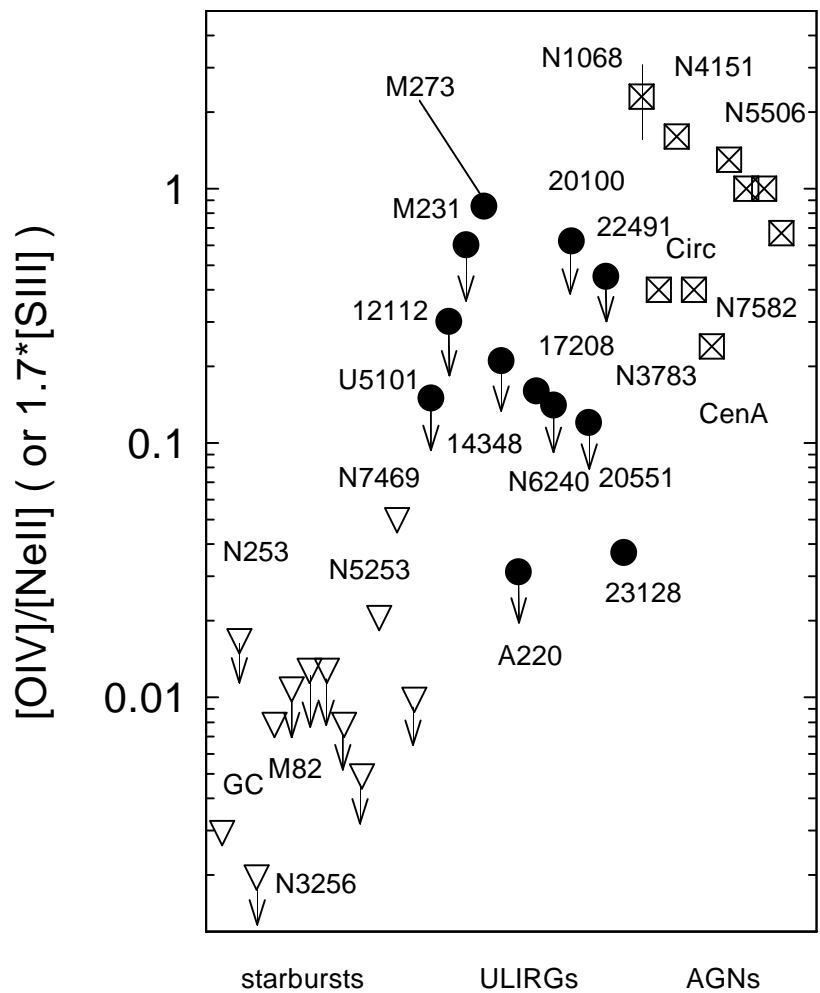
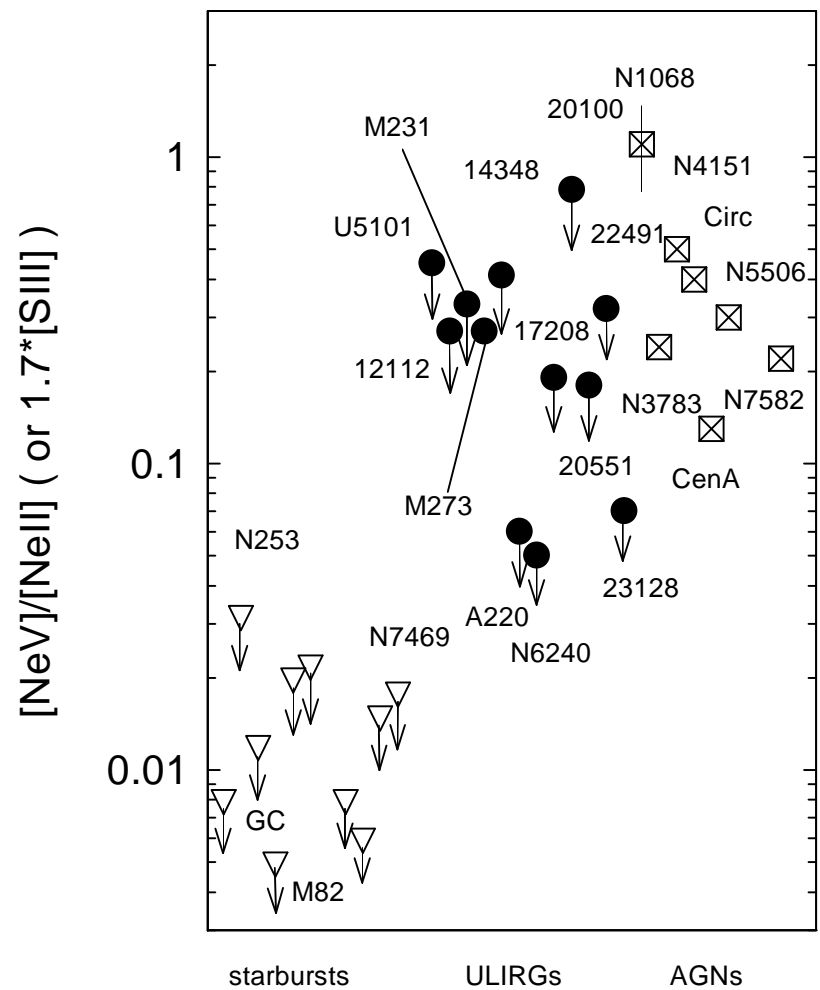

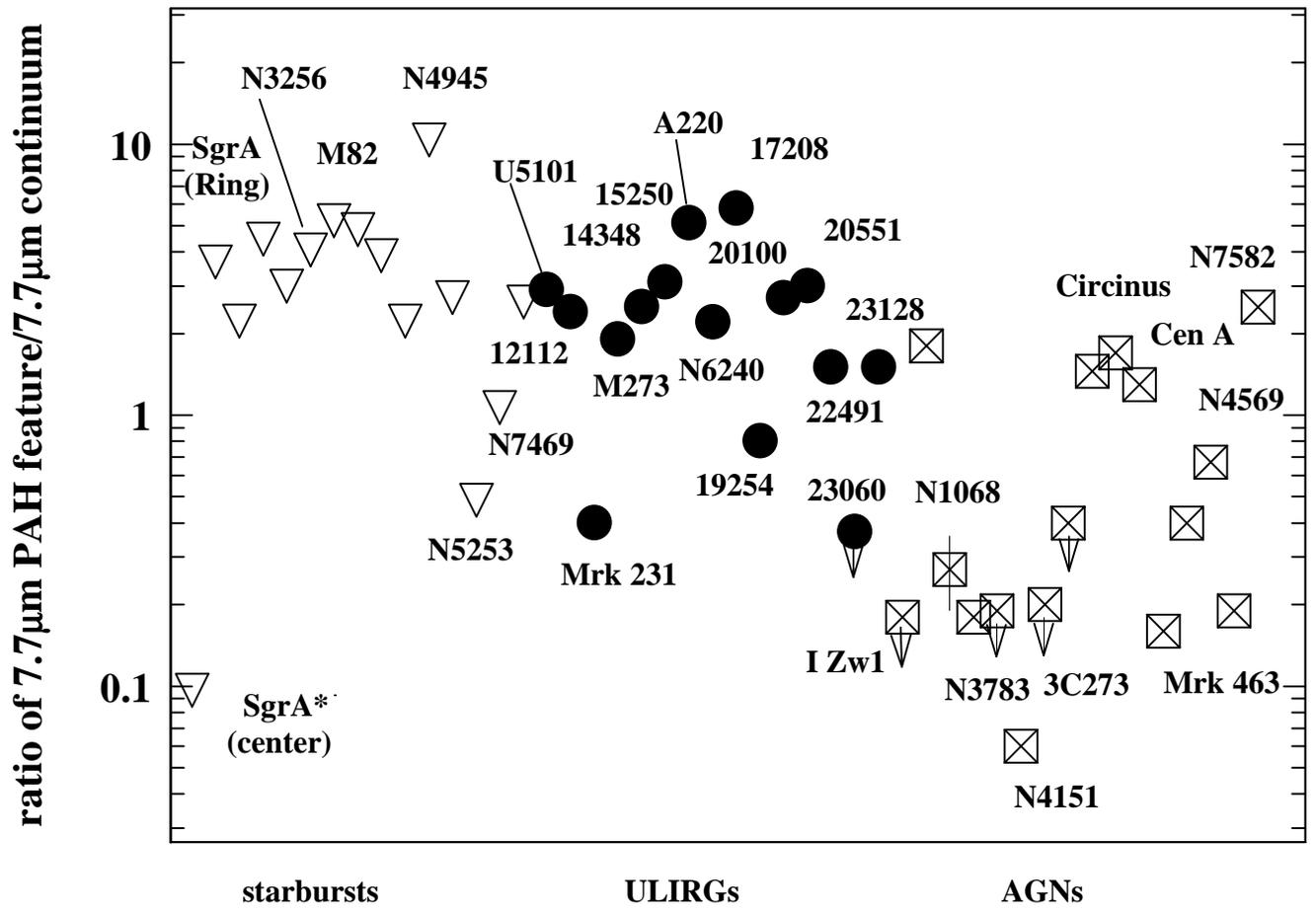

ratio of 7.7μm PAH feature/7.7μm continuum

N3256   N4945

SgrA
(Ring)   M82

U5101   A220   17208

15250

14348   20100   20551

12112   N6240   23128

M273   22491

N7469

N5253   19254   23060   N1068   Circinus   N7582

Cen A

N4569

Mrk 231   I Zw1   3C273   Mrk 463

N3783   N4151

SgrA*'
(center)

0.1

starbursts        ULIRGs        AGNs

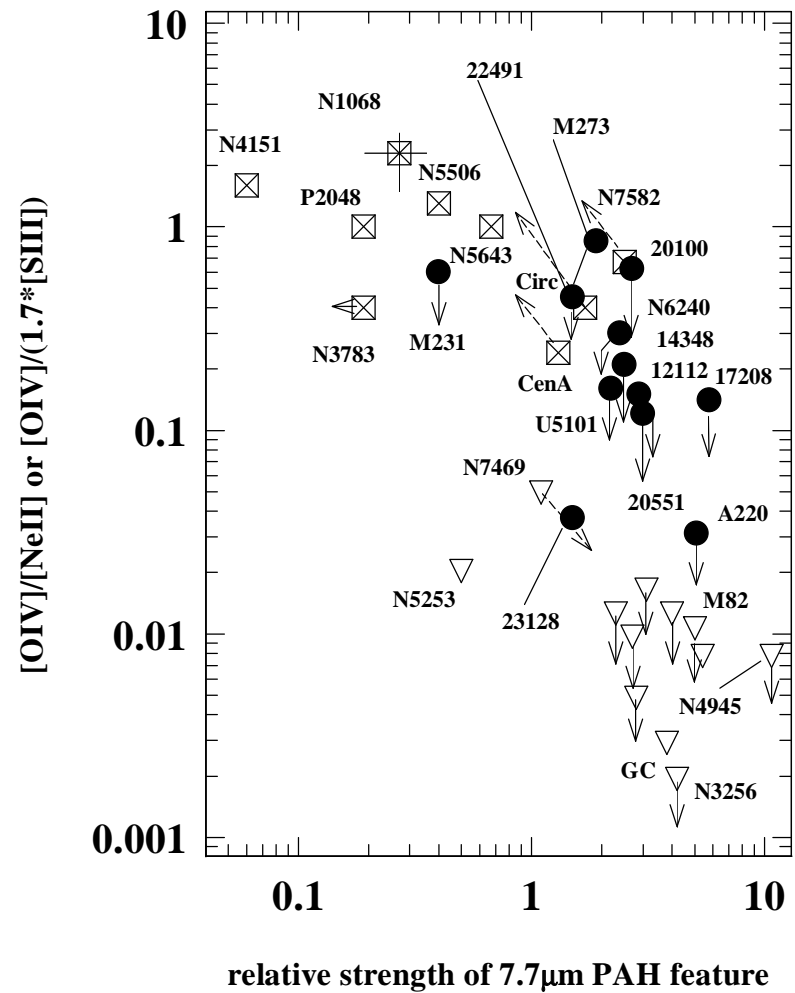
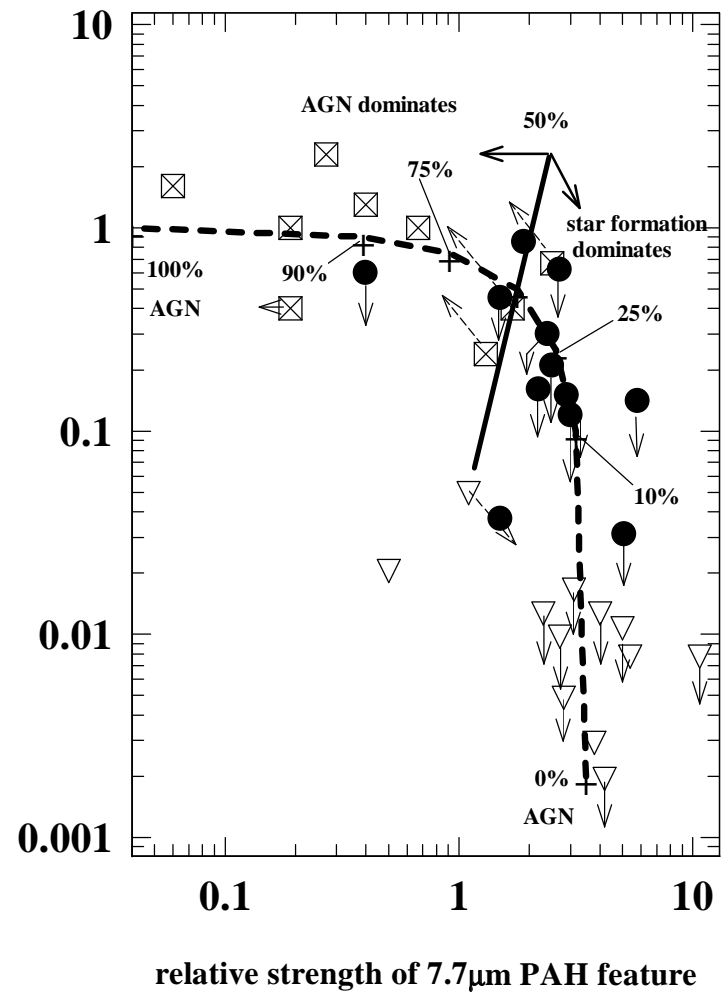

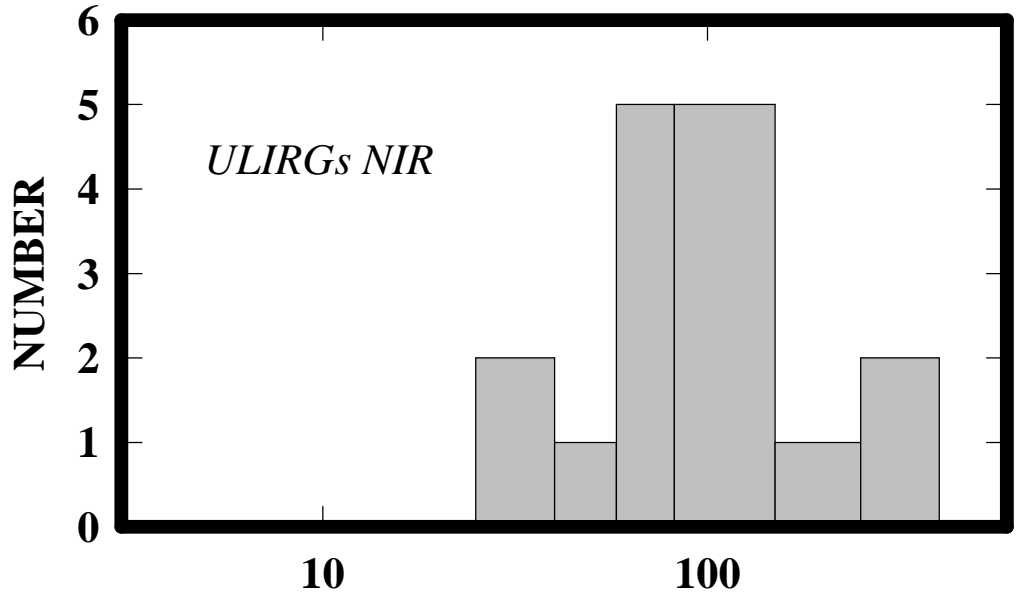

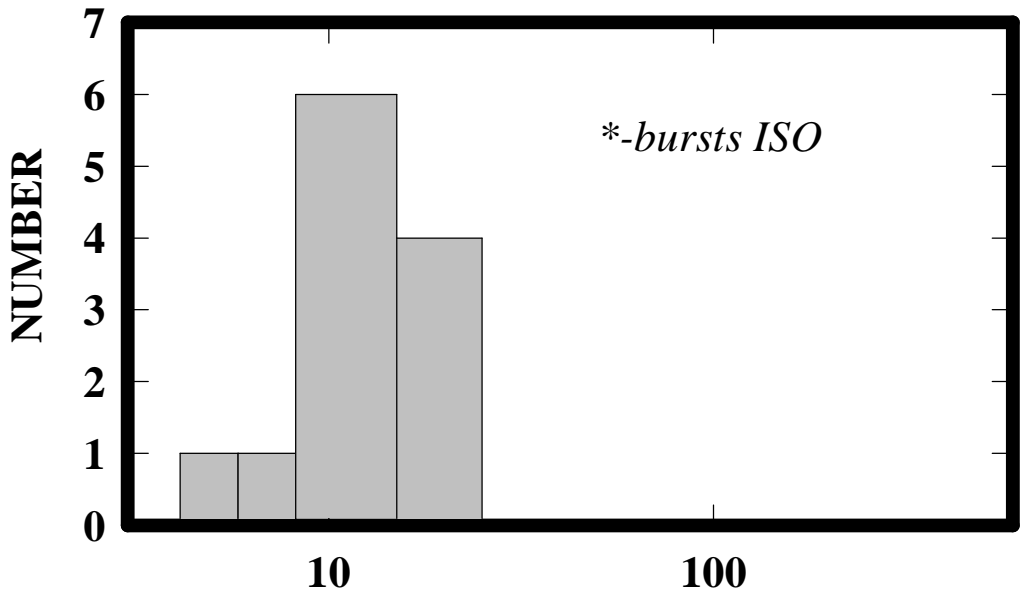

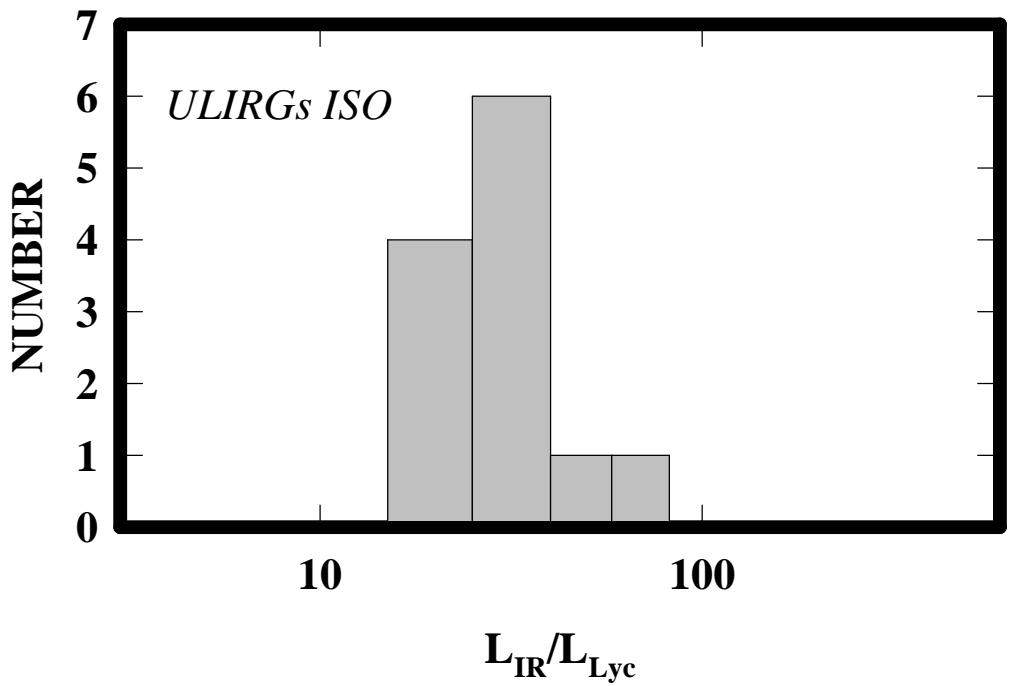

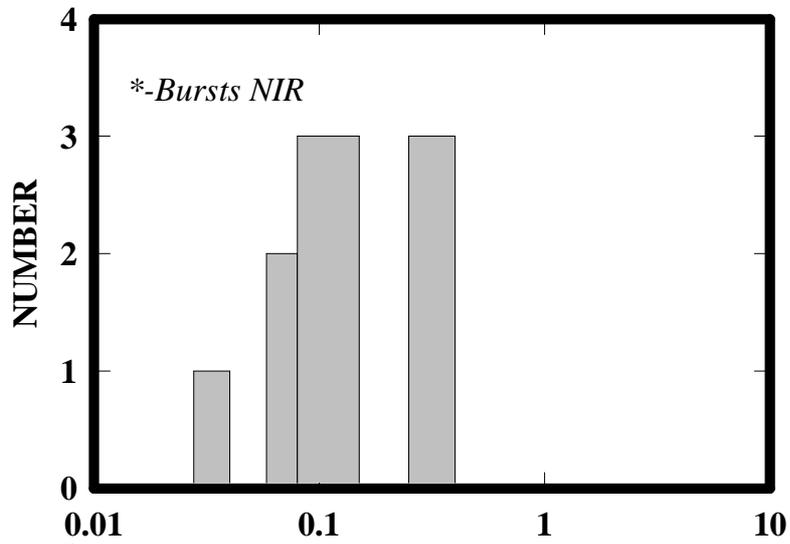

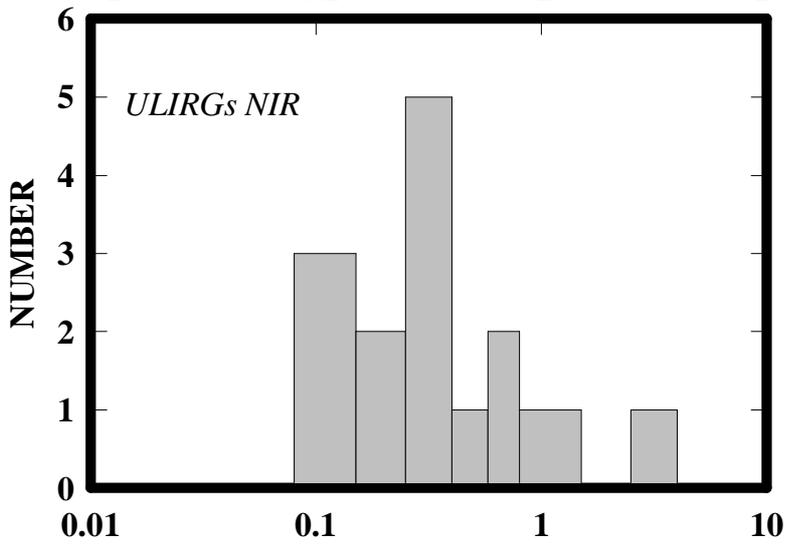

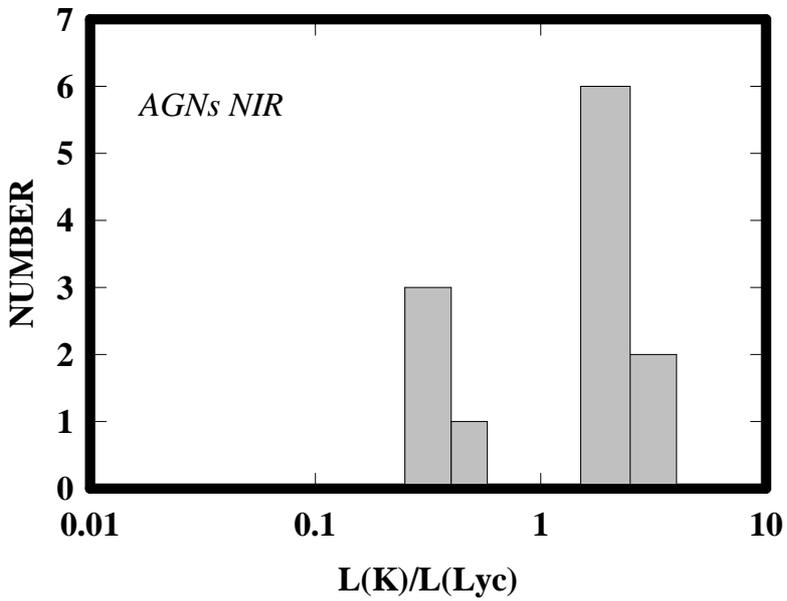

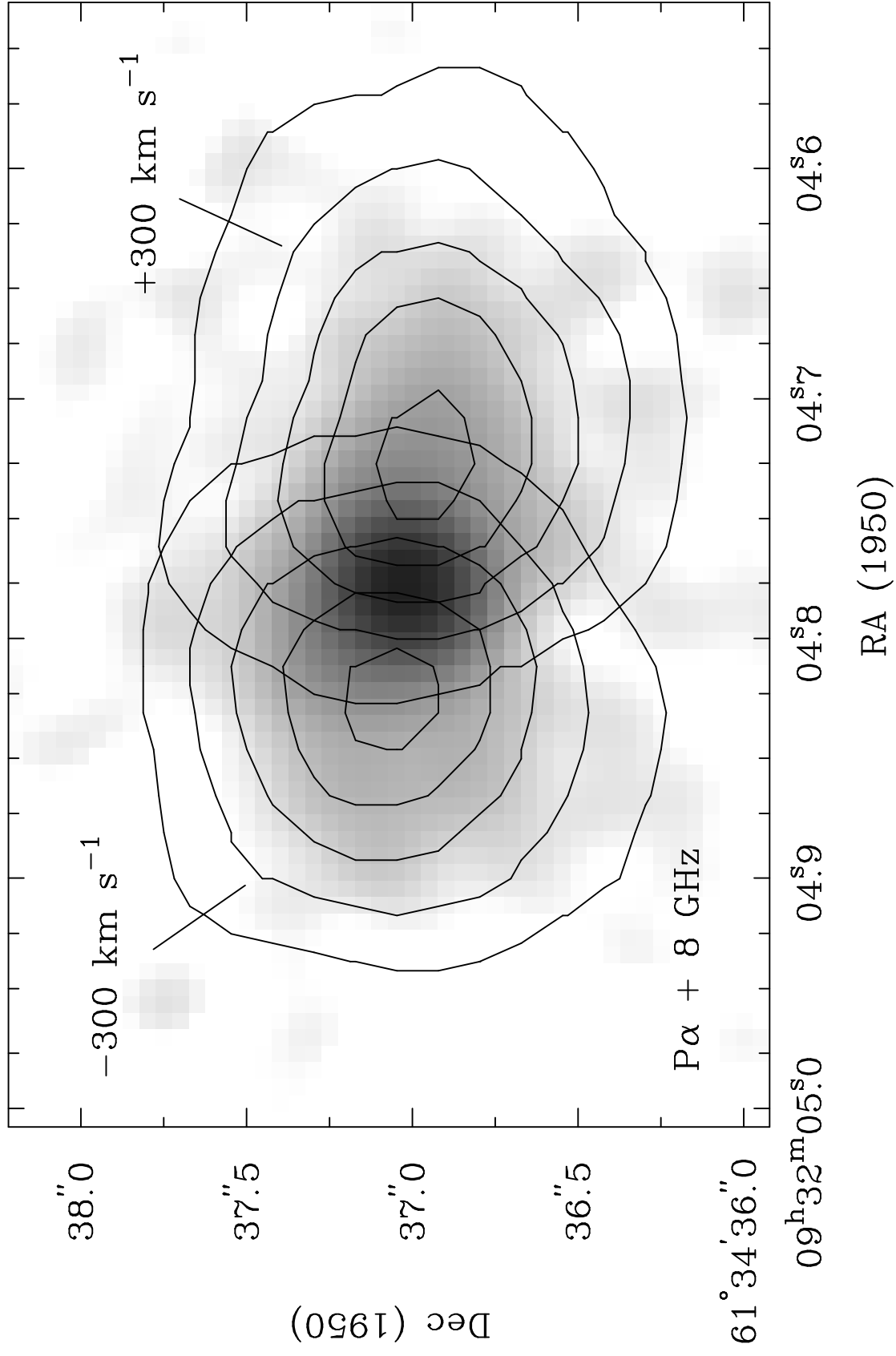

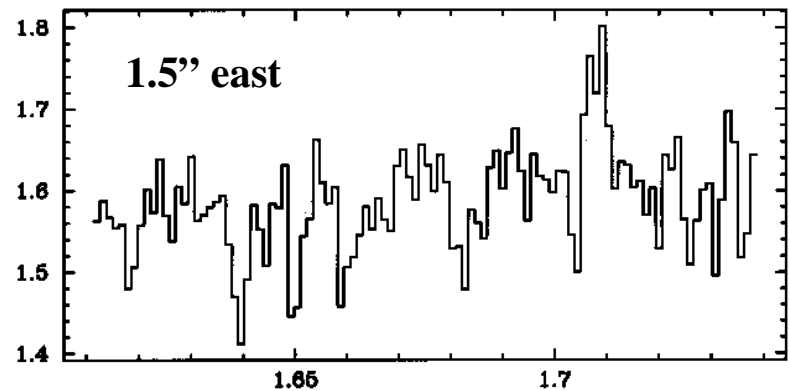

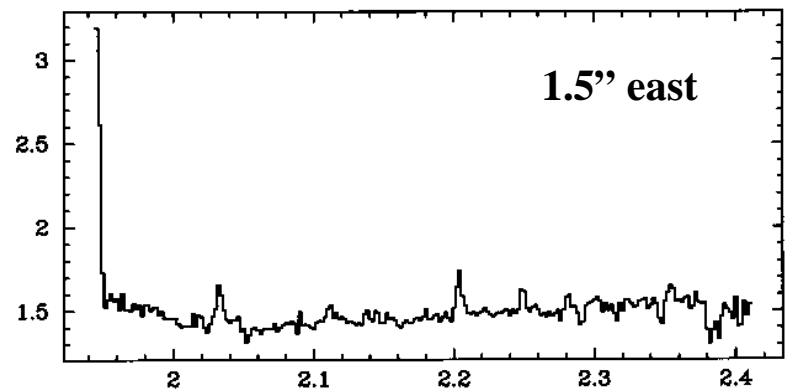

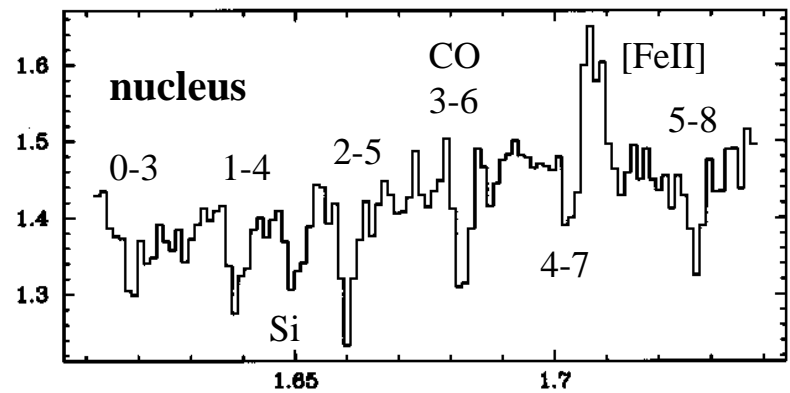

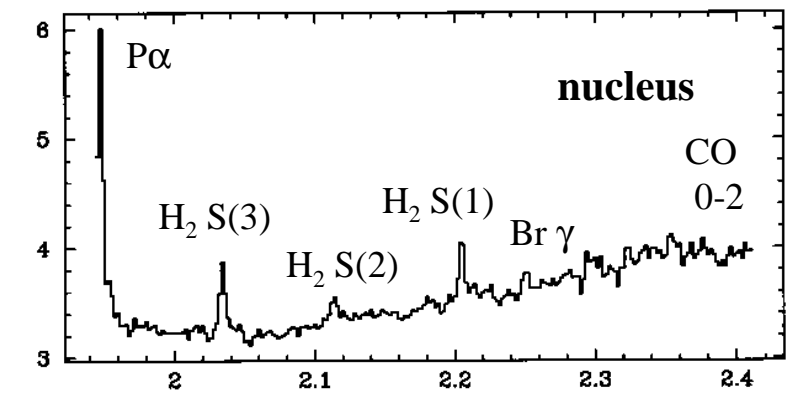

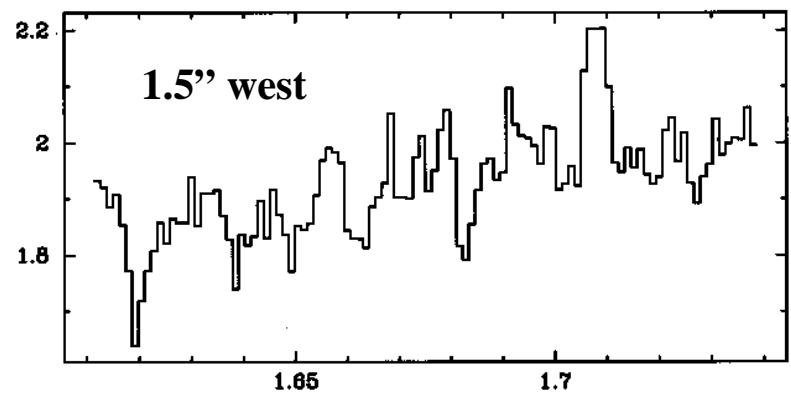

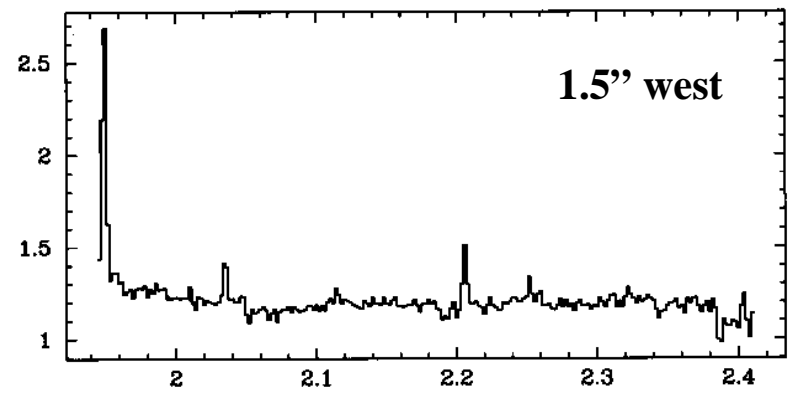

**relative flux density**

**wavelength (μm)**

In the panels (left column, top to bottom): "1.5" east", "nucleus" with line labels 0-3, 1-4, 2-5, CO 3-6, 4-7, [FeII] 5-8, Si, and "1.5" west".

In the panels (right column, top to bottom): "1.5" east", "nucleus" with line labels Pα, H$_2$ S(3), H$_2$ S(2), H$_2$ S(1), Br γ, CO 0-2, and "1.5" west".

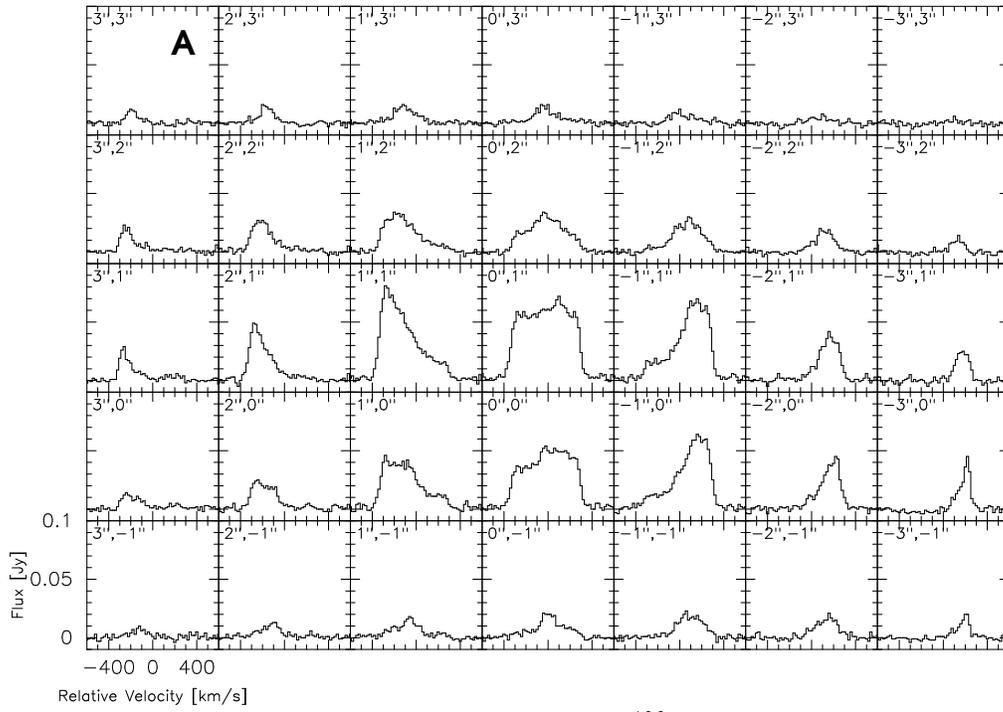

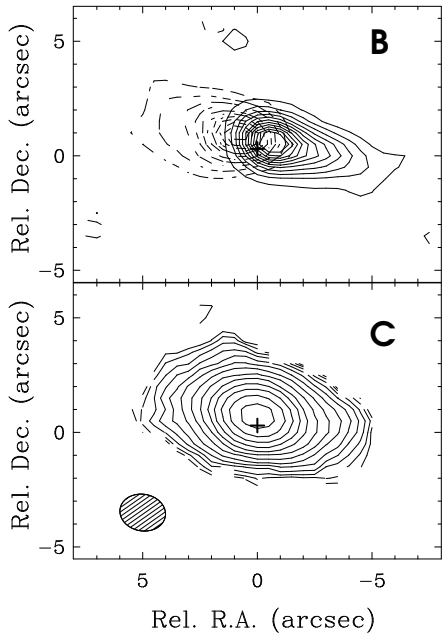

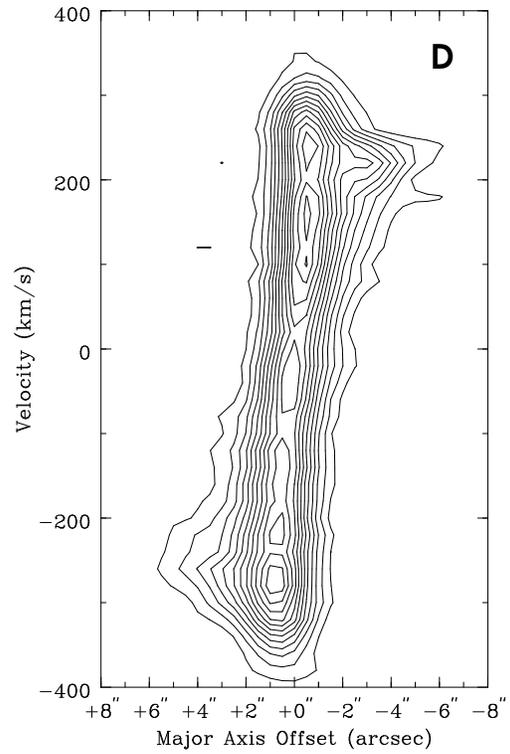

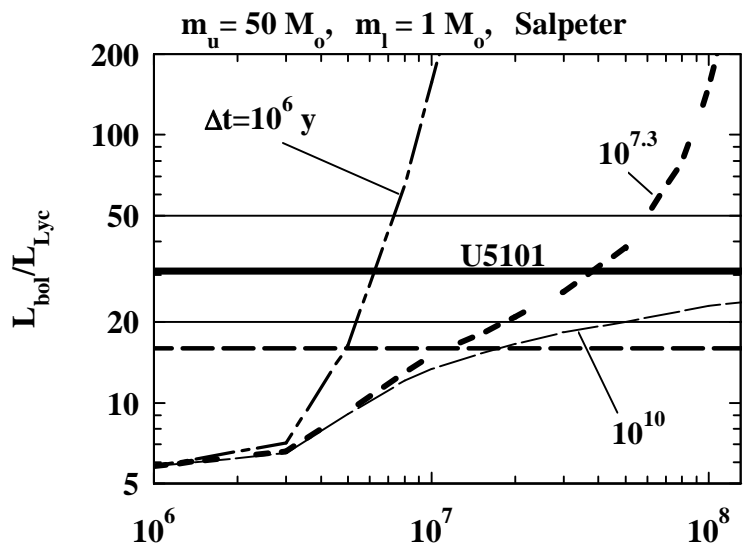

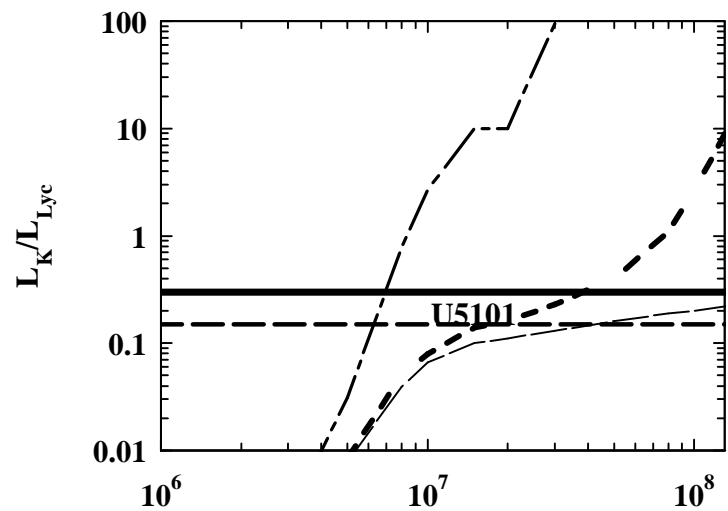

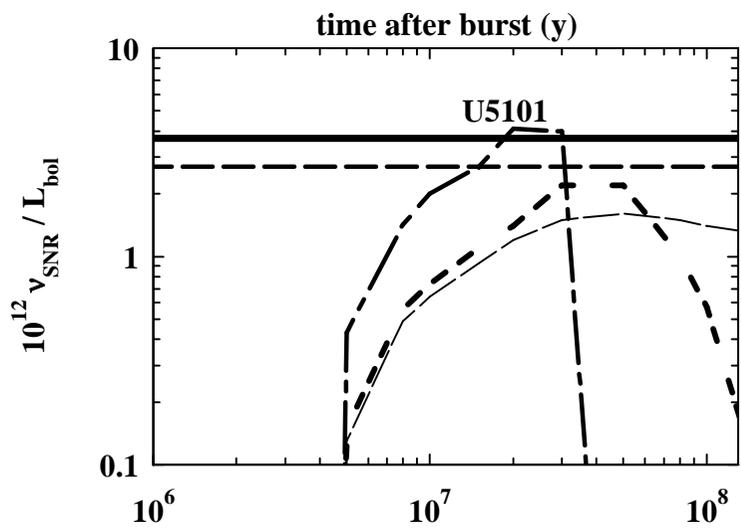

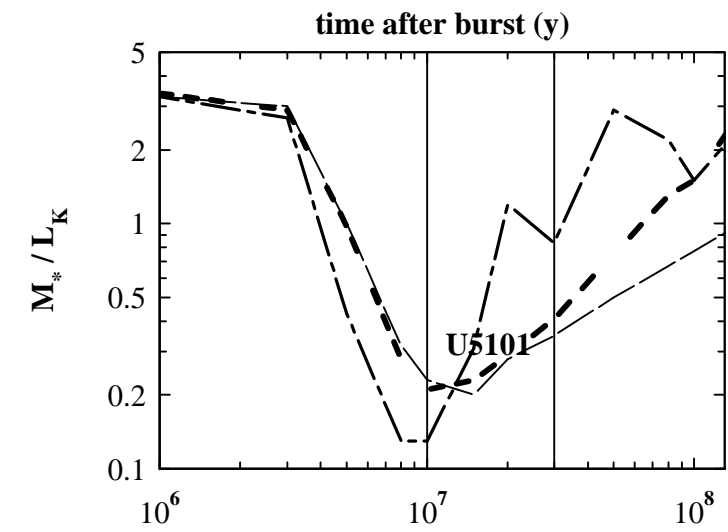

# WHAT POWERS ULTRA-LUMINOUS IRAS GALAXIES ?[1]

**R.Genzel[(1)], D.Lutz[(1)], E.Sturm[(1)], E.Egami[(1)], D.Kunze[(1)], A.F.M.Moorwood[(2)], D.Rigopoulou[(1)], H.W.W.Spoon[(1)], A.Sternberg[(3)], L.E. Tacconi-Garman[(1)], L.Tacconi[(1)] and N.Thatte[(1)]**

**(1)** *Max-Planck Institut für extraterrestrische Physik, Postfach 1603, D85740 Garching, FRG*

**(2)** *European Southern Observatory,Karl-Schwarzschildstr.2, D85748 Garching, FRG*

**(3)** *School of Physics and Astronomy, Tel Aviv University, TelAviv 68878, Israel*

## ABSTRACT

We present an ISO SWS and ISOPHOT-S, mid-infrared spectroscopic survey of fifteen ultra-luminous IRAS galaxies ($L_{IR} \geq 10^{12}$ $L_o$). We combine the survey results with a detailed case study, based on arcsecond resolution, near-IR and mm imaging spectroscopy, of one of the sample galaxies (UGC 5101). We compare the near- and mid-IR characteristics of these ultra-luminous galaxies to ISO and literature data of thirty starburst and active galactic nuclei (AGN), template galaxies. We find that

1)      70 to 80% of the ultra-luminous IRAS galaxies in our sample are predominantly powered by recently formed massive stars. 20 to 30% are powered by a central AGN. These conclusions are based on a new infrared 'diagnostic diagram' involving the ratio of high to low excitation mid-IR emission lines on the one hand, and on the strength of the 7.7μm PAH feature on the other hand.

2)      at least half of the sources probably have simultaneously an active nucleus and starburst activity in a 1-2 kpc diameter circum-nuclear disk/ring.

3)      the mid-infrared emitting regions are highly obscured ($A_V$(screen)~5 to 50 or $A_V$(mixed)~50 to 1000). In a model where star forming regions and dense molecular clouds are fully mixed the ISO derived, V-band dust extinctions approach the dust column densities inferred from CO mm measurements. After correction for these extinctions, we estimate that the star forming regions in ULIRGs have ages between $10^7$ and $10^8$ years, similar to but somewhat larger than those found in lower luminosity starburst galaxies.

[1] Based in part on observations with ISO, an ESA project with instruments funded by ESA Member States (especially the PI countries: France, Germany, the Netherlands and the United Kingdom) with the participation of ISAS and NASA.

4)      in the sample we have studied there is no obvious trend for the AGN component to dominate in the most compact, and thus most advanced mergers. Instead, at any given time during the merger evolution, the time dependent compression of the circum-nuclear interstellar gas, the accretion rate onto the central black hole and the associated radiation efficiency may determine whether star formation or AGN activity dominates the luminosity of the system.

Key words: galaxies: stellar content - galaxies: starburst - galaxies: nuclei - infrared: sources - infrared: galaxies- stars: formation

# 1. INTRODUCTION

Many different studies have established that ultra-luminous infrared galaxies (ULIRGs : L(8-1000μm) $\geq 10^{12}$ L$_o$, Soifer et al. 1984, 1987) are gas rich galaxies that in most cases have undergone a recent strong interaction with another galaxy eventually leading to a complete merger of the two (for recent reviews see Sanders and Mirabel 1996 and Moorwood 1996). A commonly accepted interpretation is that during such a galaxy collision the interstellar media experience strong, large scale torques and loss of angular momentum. As a consequence interstellar gas is transported toward the circum-nuclear environment, is concentrated and compressed there, thus resulting in prodigious star formation on scales of less than a few kiloparsec (e.g. Norman 1991, Barnes and Hernquist 1996). In a next phase very low angular momentum gas produced in part by stellar mass loss in the starburst zone may fall into the nucleus proper and accrete onto the central massive black hole (Norman and Scoville 1988, Scoville and Norman 1988). Luminosities and space densities of ULIRGs in the local Universe are similar to those of quasi-stellar objects (QSOs) (Soifer et al. 1987). Based on optical, near-infrared and millimeter photometry and spectroscopy Sanders et al.(1988a) have proposed that most ULIRGs are predominantly powered by dust enshrouded QSOs in the late phases of such a merger process.

Despite this fairly plausible scenario we have just sketched, the central questions of *what dominates on average* the luminosity of ULIRGs and *how ULIRGs evolve* are by no means answered. On the one hand the infrared, millimeter and radio characteristics of ULIRGs are very similar to those of starburst galaxies (Rieke et al. 1985, Rowan-Robinson and Crawford 1989, Condon et al. 1991b, Rigopoulou et al. 1996, Goldader et al. 1995,1997a,b). Their extended optical emission line nebulae resemble the expanding 'superwind bubbles' of powerful starburst galaxies (Armus et al. 1990, Heckman et al. 1990). On the other hand, a significant fraction of ULIRGs exhibit nuclear optical emission line spectra characteristic of Seyfert galaxies (Sanders et al. 1988a, Armus et al. 1990, Kim et al. 1996, Veilleux et al. 1995, 1997). Some contain compact central radio sources (Lonsdale et al. 1993, Lonsdale, Smith and Lonsdale 1995) and highly absorbed, hard X-ray sources (Mitsuda 1995, Brandt et al. 1997, Kii et al. 1997), all indicative of an active nucleus (AGN). ***Adopting henceforth the working asumption that a central AGN as well as circum-nuclear star formation may both play a role, the general questions we want to pursue in the present paper are which (if any) of the two dominates on average and whether there is a trend with merger evolution.***

    With the advent of the Infrared Space Observatory (ISO) of the European Space Agency (Kessler et al. 1996) on the one hand, and with recent improvements in sensitive, arcsecond resolution near-infrared and millimeter imaging spectroscopy on the other hand, it is now possible to take a fresh look at the issues of the energetics, dynamics and evolution of luminous infrared galaxies. The following article is structured as follows. After a description of the observations and data reduction in section 2, we describe in section 3 our ISO global spectroscopic survey of a sample of fifteen ULIRGs and the statistical conclusions that can be drawn from that. We compare the results to ISO observations of about thirty starburst and AGN templates, as well as other relevant data in the literature. Our new results follow up on our first ISO papers on the nature of ultra-luminous galaxies (Lutz et al. 1996a, Sturm et al. 1996) and put that work on a more secure statistical footing. To test several key predictions of the survey, we take in section 4 the 'case study' approach and report on detailed spatially resolved, multi-wavelength imaging spectroscopy of one of the galaxies in the sample, and summarize similar observations for a second one. In the last two sections (5,6) we discuss the implications of our observations for the starburst/AGN connnection and the evolution of ultra-luminous galaxies.

# 2. OBSERVATIONS AND DATA REDUCTION

## 2.1 ISO SWS and ISOPHOT-S spectroscopy

As part of the MPE ISO Central Program project 'galactic nuclei', we observed between January and December 1996 thirteen ULIRGs with L(8-1000μm)[2] ~ $10^{11.8}$ to $10^{12.5}$ $L_o$ with the short-wavelength grating spectrometer (SWS: deGraauw et al. 1996) onboard ISO. To this we have added NGC 6240 which, although it is slightly less luminous than the others, exhibits all of the ULIRG characteristics. In each source between 6 and 24 different spectral lines were observed in the AOT SWS02 mode (grating line profiles scans) with a resolving power $\lambda/\Delta\lambda$~1500, for a total integration per source between 2 and 10 hours. IRAS 19254-7245 was observed in SWS06 mode, giving larger spectral ranges around each line, because optical spectroscopy (Mirabel, Lutz and Maza 1991) shows ~1700 km/s widths of the forbidden lines. The ISO SWS aperture sizes range between 14"x20" and 20"x33" so that we can safely assume that essentially all of the mid-infrared line flux is contained in it.

The SWS data were reduced with the Interactive Analysis (IA) Package of the SWS team (see deGraauw et al.1996) putting special emphasis on tools to improve cosmic ray spike removal, dark subtraction and flat fielding. The final spectra are oversampled with respect to the binning size of typically ~200km/s which we used to combine the many individual scans. The absolute flux scale of SWS is accurate to about ±25%. There is a larger uncertainty for the continuum and its slope than for the line fluxes, since some of our observations were obtained early in the mission before optimization of the SWS dark current measurement strategies. Also, the continuum contains a small contribution by zodiacal light. Error sources contributing to the `noise' in the final spectra include not only detector noise and after effects of ionizing particle hits, but also systematic effects like signal drifts and residual fringing. Instead of relying on error propagation, we have measured the noise in all individual final spectra in regions free of line emission, and indicated the result in Fig. 1. The achieved sensitivity varies significantly due to different integration times and because some observations were obtained with preliminary observing strategies. We achieve minimum detectable line fluxes (with good confidence) of typically a few $10^{-21}$ W cm$^{-2}$. Wherever we quote upper limits in Table 1, the values given are such 'most probable' flux limits. The specifics of the SWS observations of several of the brightest ULIRGs are reported in several other papers (Arp 220: Sturm et al. 1996, 1997, Mrk 231: Rigopoulou et al. 1997, NGC 6240: Egami et al. 1997). Some of the new results on the starburst galaxies and AGNs are also already published or have been submitted for publication, some are awaiting publication. Details can be found in Moorwood et al.(1997: NGC 4945), Lutz et al. (1996b: Galactic Center, 1997: NGC 1068), Kunze et al.(1996: NGC 4038/39, 1997: NGC 5253), Förster-Schreiber et al. (1997: M82), Moorwood et al. (1996: Circinus), Rigopoulou et al. (1996: NGC 3256), Alexander et al. (1998: NGC 4151).

---

[2] For calculating luminosities etc. we are using throughout this paper distances based on $H_o$=75 km/s/Mpc and $q_o$=0.5. For deriving L(8-1000μm) from the IRAS fluxes we employ the formula for $L_{IR}$ given in Table 1 of Sanders and Mirabel (1996, see also footnote in Table 2).

3-11.6μm spectra of the same ULIRGs that were observed with SWS (plus one additional source) were also obtained at lower spectral resolution ($\lambda/\Delta\lambda$~90) with the ISOPHOT-S spectrophotometer (Lemke et al. 1996). ISOPHOT-S comprises two low resolution grating spectrometers simultaneously covering the wavelength bands 2.47 to 4.87μm and 5.84 to 11.62μm. The spectrum is registered by two linear arrays of 64 Si:Ga detectors with a common entrance aperture of 24"x24". Observations of our ULIRG sample (plus other galaxies described below) were done in the chopped mode, using a pure on-source integration time of 512 seconds. The resulting spectra thus are free of a contribution from zodiacal light that otherwise may affect faint sources. A few bright sources in the Galaxy, as well as the Circinus galaxy and NGC 253 were observed in staring mode and shorter integration times. The data processing was performed using standard procedures of the PHT Interactive Analysis (PIA) software[3], version 6.0. PIA also computes an error estimate which is propagated into the final spectrum. This error estimate is illustrated in Fig.2 by adding the corresponding error bars to the spectrum of UGC5101. The PIA error estimate reasonably reproduces the effect of detector noise on the ISOPHOT-S spectra, most prominently the significant increase of noise at wavelengths beyond 9μm. However, it does only partially cover the effects of slow signal drifts which are known to be important for low background photoconductors as used by ISOPHOT. Drifts are similar for neighbouring pixels. At significant source flux, the formal PIA error returned for a certain pixel is much higher than the actual uncertainty of detecting a spectral feature, by comparison to the neighbouring pixels. On the other hand, the PIA error underestimates the drift-induced uncertainty in the global flux scale of an ISOPHOT-S spectrum. Recent improvements in the modeling of detector drifts for chopped ISOPHOT-S data indicate that flux densities of chopped observations like ours have to be increased by ~40% compared to results without drift handling, as in PIA version 6.0 (U.Klaas, priv. comm.). This is however mostly a scaling of the entire spectrum, and uncertainties of feature-to-continuum ratios as derived below from our spectra will be less than 30%.

## *2.2 3D Imaging Spectroscopy*

UGC 5101 (IRAS 09320+6134, MGC +10-14-025, cz=11,800 km/s, distance=158 Mpc, 1"=770 pc) was observed with the MPE near-infrared field imaging spectrometer 3D (Weitzel et al. 1996) in January 1995 on the 3.5 m telescope on Calar Alto (Spain) and in January 1996 on the 4.2 m William Herschel telescope (WHT) on La Palma (Spain). For detailed descriptions of the 3D instrument and data reduction we refer to Krabbe et al. (1995) and Weitzel et al.(1996). Briefly, 3D records 256

---

[3] PIA is a joint development by the ESA Astrophysics Division and the ISPOHOT Consortium led by the Max-Planck Institut für Astronomie (MPIA), Heidelberg. Contributing ISPOHOT Consortium institutes are DIAS, RAL, AIP, MPIK and

channel spectra simultaneously for each pixel of a contiguous 16x16 pixel square field on the sky. For the Calar Alto observations 3D was set up to cover the entire K-band (1.95 to 2.4μm) at resolving power $\lambda/\Delta\lambda$=1000 per pixel and 0.5" per pixel. The seeing was below 1" so that the K-band data had effectively 0.9" FWHM resolution. For the La Palma observations 3D was set up to cover the H-band (1.4-1.8μm) at resolving power 1000 per pixel and 0.3" per pixel. Again the seeing was sub-arcsecond so that in combination with the fast tip-tilt guider ROGUE in front of 3D (Thatte et al. 1995) the data have a FWHM resolution of 0.75".

On-source exposures (~60 to 100 secs) were interleaved with an equal amount of time spent ~1' off the source for sky subtraction. The total on-source integration time was 2000 seconds for the K-band data and 2800 seconds for the H-band data. The data reduction was carried out with software developed at MPE (Krabbe and Thatte, unpublished) in the GIPSY (van der Hulst et al. 1992) spectral line data reduction environment. Individual frames were first corrected for non-linear detector response, then sky subtracted and spatially flat fielded with a dome exposure. The spectral calibration with a Neon discharge lamp included rebinning of the frames onto a 300 channel linear wavelength grid. For a spectral flatfield, we used a lamp that produces an almost flat near-infrared spectrum. After rearranging the data into a 16x16x300 data cube, bad pixels and cosmic ray hits were removed and interpolated, or blanked out in the corrresponding 'mask' cube. Before coadding the data cubes individual frames were rebinned onto a 0.25" (0.15" for H-band) grid and then recentered on the nuclear source. Telluric absorption was then corrected for by dividing the cubes by an (almost featureless) late F/early G dwarf star in the case of the K-band data, and by an A star spectrum in the case of the H-band data. In both cases stellar hydrogen absorption features were removed by interpolation. While care was taken to pick stars at very similar airmass to the galaxy, a scaling of the observed telluric features to exactly the airmass of UGC 5101 was obtained with the ATRAN software package (Lord 1992). For absolute flux calibration we adopted the aperture fluxes of Sanders et al. (1988a).

## 2.3 CO 1-0 Millimeter Interferometry

We mapped the 2.6mm $^{12}$CO J=1-0 line emission from UGC 5101 between September 1996 and February 1997 with the IRAM millimeter interferometer on Plateau de Bure, France (Guilloteau et al. 1992). The 5 15-meter antennas were positioned in 4 different configurations, for a total of 36 baselines ranging from 24 to 288 meters. We observed UGC 5101 for ~8 hours in each configuration. The antennas were equipped with SIS receivers with single sideband (SSB) system temperatures (including all telescope and atmospheric losses) of ~170 K. We observed 3C 273 and 0415+379 for bandpass calibration. In addition, we observed 0923+392 or 1044+719 every half hour to calibrate the phases and amplitudes. Absolute fluxes were established by assigning 111 GHz flux densities of 6.5 and 26 Jy for 0415+379 and 3C 273, respectively. Fluxes of strong sources are determined through careful monitoring with both the interferometer and the IRAM 30-meter telescope on Pico de Veleta, Spain. Based on the monitoring measurements, we conclude that the accuracy of the flux density scale is better than 30%. The phase noise on the longest baselines was $20^o$ to $30^o$. Spectral resolution of 2.5 MHz (6.5 km/s) was provided by 5 autocorrelator spectrometers covering the total receiver bandwidth of 5000 MHz (1300 km/s) centered on z=0.0395 (cz=11,840 km/s). The data were first calibrated with the IRAM CLIC software. We then made uniformly weighted channel maps which we CLEANed using software available as part of the GILDAS reduction package. The maps were made with a velocity resolution of 20 km/s and a FWHM spatial resolution of 2.1"x1.6" (p.a. $80^o$ east of north). The rms noise in the CLEANed maps is 2 mJy/beam.

# 3. MID-IR SPECTROSCOPIC SURVEY WITH ISO SWS AND ISOPHOT-S

In this section we discuss the results of the ISO mid-infrared spectroscopic survey of a sample of ULIRGs. Figure 1 shows the key mid-infrared/far-infrared lines we have observed thus far with SWS in fourteen ULIRGs: 12.8μm [NeII], 14.3μm [NeV], 18.7 and 33.5μm [SIII], 25.9μm [OIV] and 34.8μm [SiII]. Figure 2 shows the ISOPHOT-S 5.8-11.6μm spectra of all fifteen ULIRGs we have observed. Of these, eight are from the IRAS Bright Galaxy Sample (Sanders et al. 1988a: S(60μm)≥5.4 Jy, │b│≥30$^o$, Dec.≥-30$^o$) and five are southern ULIRGs listed in the southern extension of the IRAS bright galaxy survey (Sanders et al. 1995). IRAS 23060+0505 is the most distant source in our sample (z=0.17, cz=51,900 km/s) which is a S(60μm)~1 Jy, broad-line ULIRG (see Veilleux et al. 1997). As already mentioned above we have added to this list the advanced merger NGC 6240 which, although it is slightly less luminous than the others, exhibits all of the ULIRG characteristics. The basic ISO SWS and ISOPHOT-S data for the ULIRGs are given in *Table 1.*

*Table 2* summarizes other basic and deduced properties of all fifteen ULIRGs, along with 27 other infrared bright galaxies, plus the central parsec of our own Galaxy. The table is split into 12 galaxies that are known to be (predominantly) powered by star formation ('STARBURST TEMPLATES'), 16 galaxies predominantly powered by AGNs and the ULIRGs (including UGC 5101). Among the templates there are several known composite sources with both AGN and starburst activity. NGC 7469 has a Seyfert 1 nucleus surrounded by a kpc ring of star formation. Genzel et al. (1995) have shown that at least two thirds of the bolometric luminosity comes from the starburst ring. Moorwood et al.(1996) have estimated that about half of the bolometric luminosity of the Circinus galaxy comes from the Seyfert 2 nucleus and the other half from a 0.5 kpc starburst disk/ring. The nearby radio galaxy Cen A and the narrow line X-ray galaxy NGC 7582 are also composite objects.

In the following section we will explore what we can deduce from the new mid-/far-infrared spectroscopy about the source(s) of luminosity in ULIRGs with an infrared luminosity near $10^{12}$ $L_o$. Our sample contains 5 sources just under $10^{12}$ $L_o$, 6 with 1 to $2x10^{12}$ $L_o$, 3 with 2 to $3x10^{12}$ $L_o$ and 1 with greater than $3x10^{12}$ $L_o$. Those of our sources that were in the Sanders et al.(1988a) BGS sample were all listed there as having $\geq 10^{12}$ $L_o$. The small discrepancy between our and the Sanders et al. luminosities comes from the different ways Sanders et al.(1988a) and Sanders and Mirabel (1996) obtain the 8-1000µm luminosity from the IRAS fluxes.

### 3.1 Extinction from SWS Line Ratios

Whenever possible we have attempted to derive extinctions for the infrared line emitting regions from the SWS data themselves. In Arp 220 we had three hydrogen recombination lines (2.63µm Brβ, 4.05µm Brα and 7.46µm Pfα, see Sturm et al. 1996, 1997) so that a fairly accurate extinction can be obtained by comparing the observed ratios to those expected from case B recombination theory. In ten ULIRGs (including Arp 220) we have available the ratio of the two [SIII] lines ([SIII]18.7/[SIII]33.5µm) or an upper limit to it. The [SIII] 18.7/33.5µm ratio increases with electron density and is independent of the ionization parameter and the spectral shape of the UV radiation field. By adopting the low density limit for the intrinsic value of that ratio ([SIII] 18.7/33.5µm ~0.5 to 0.6 for $n_e \leq 10^{2.5}$ cm$^{-3}$) the observed line ratio then gives a **lower limit** to the 18-33µm extinction. We treated the starburst and AGN template sources in the same manner. We used the infrared extinction curve of Draine (1989), amended by the more recent ISO-SWS extinction curve toward the Galactic center for 2.5-10μm (Lutz et al. 1996b)[4]. The derived V-band extinctions are listed in columns 8 and 9 of Table 2. For Mrk 231 and IRAS 22491-1808 we adopted the extinctions enclosed by a bracket in the extinction column. For UGC 5101 we adopted an extinction which is an average of ISO and near-infrared (section 4) derived values. We distinguish between two extreme cases of dust extinction. If the obscuration is due to a foreground screen of dust with optical depth $\tau(\lambda)=0.916\ A(\lambda)$, the correction is $\exp(\tau(\lambda))$ (column 8). In the other case where emitting gas and dust are completely spatially mixed, the correction is $(\tau(\lambda)/(1-\exp(-\tau(\lambda))))$ (column 9, e.g. Genzel et al. 1995). In several cases the mixed model fits the data over a wider wavelength range (mid-infrared as well as near-infrared data) significantly better in which case that was chosen in Table 2. $A_V$(screen)~10 corresponds to $A_V$(mixed)≥100 and $A_V$(screen)~45 corresponds to $A_V$(mixed)~1000. With the exception of Arp 220, the uncertainties in these extinctions are large (50% to a factor of 2) and often, just lower limits. Yet because mid-infrared extinctions are a small fraction of the visible extinctions (screen $A_V$ ~5 to 45 corresponds to $A_{25\mu m}$ ~0.1 to 0.9) and because most of the scientific conclusions below depend on line ***ratios***, our results are fairly insensitive to the adopted extinction values and dust distributions.

### *3.2 Infrared Fine Structure Lines as Probes of the UV Radiation Field*

In the 33 sources of Table 2 with SWS data, we have observed lines from ions with a wide range of excitation potential, ranging from 8.2 eV for $Si^+$, 13.6 eV for the hydrogen recombination lines, 21 to 41 eV for $Ne^+$ and $S^{++}$, to 55 eV for $O^{+++}$ and 97 eV for $Ne^{4+}$. With the exception of $Si^+$, these lines arise in HII regions that are photoionized by stars or a central active nucleus (e.g. Spinoglio and

---

[4] The ratios A(λ)/A(V) at important wavelengths in the ISO-SWS range then are as follows: 0.1 at 2.17μm (Brγ), 0.077 at 2.63μm (Brβ), 0.043 at 4.05μm (Brα), 0.049 at 7.46μm (Pfα), 0.04 at 12.8μm ([NeII]), 0.038 at 14.3μm ([NeV]), 0.044 at 18.7μm ([SIII]), 0.02 at 25.9μm ([OIV]) and 0.012 at 33-35μm ([SIII],[SiII]).

Malkan 1992, Voit 1992). Some contribution from fast, ionizing shocks is possible (e.g. Contini and Viegas 1992, Sutherland, Bicknell and Dopita 1993). The [SiII] 34.8μm emission originates in HII regions and/or in dense photon-dominated regions (PDRs) at the interfaces between HII regions and adjacent dense molecular clouds (e.g. Tielens and Hollenbach 1985, van Dishoeck and Black 1988, Sternberg and Dalgarno 1989). Electron densities can be obtained from [SIII] 18.7/33.5μm (or from the [NeIII] 36.0/15.6 μm) and -after correction for extinction (see above) - yield fairly constant electron densities ($n_e \sim 10^2$ to $10^{2.5}$ cm$^{-3}$) in the galaxies we have observed so far (Lutz et al. 1996a, Rigopoulou et al. 1996, Kunze et al. 1996, Förster-Schreiber et al. 1997). High to low excitation ionic line ratios thus are a tool for probing the hardness (effective temperature) of the ionizing UV radiation field. Line ratios from different ions of a single atom are preferable, such as [NeV] 14.3/[NeII] 12.8μm. In addition we use below the ratio [OIV] 25.9/[NeII] 12.8μm. Whenever the [NeII] line is not available we take instead [SIII] 33.5μm. Model calculations with the photoionization code CLOUDY (Ferland 1993) show that for a range of ionization parameters (log U~-2.5 to -1.5) and effective temperatures (31,000 to 42,000 K) the ratio of [NeII] to [SIII] should be fairly constant if their abundance ratio is constant. This result does not depend on the stellar atmosphere models used, either the line-blanketed LTE models of Kurucz (1992), or the new line-blanketed NLTE models of Pauldrach et al. (1997) (including stellar winds). In this parameter regime both lines should closely track the Lyman continuum luminosity. In the starburst galaxies we find empirically that the [NeII] line is about 1.7 times brighter than the [SIII] 33.5μm line.

### 3.3 High Excitation (Coronal) Lines : Finding ULIRGs Dominated by AGNs

Fig.3 shows the [OIV]/[NeII] (or [OIV]/1.7 [SIII]) and [NeV]/[NeII] (or [NeV]/1.7 [SIII]) ratios for the star formation powered sources, for the AGNs and the ULIRGs, all taken from Table

2. Four of twelve galaxies powered by stars show detectable 25.9μm [OIV] emission but in all (the Galactic center, M82, NGC 5253 and NGC 7469) the lines are much weaker than in the 9 AGNs. NGC 7469 is a composite source with a prominent Seyfert 1 nucleus, in addition to a starburst ring that dominates the luminosity (Genzel et al. 1995). NGC 5253 is a low metallicity, blue compact dwarf galaxy. Dwarf galaxies are well known to have highly excited HII regions (Kunze et al. 1997, see also Moorwood 1996 for a general discussion). Among the ULIRGs Mrk 273 has a strong and clearly detected [OIV] line. NGC 6240 and IRAS 23128 have weaker [OIV] lines, probably indicating the presence of an AGN. The [NeV] data support the [OIV] results but are generally of lower significance. While none of 12 galaxies powered by star formation show [NeV], all 9 AGNs do. Only 1 of the 14 observed ULIRGs (Mrk 273) shows a [NeV] detection at the ~4σ level. For both [NeV] and [OIV] the case of Mrk 231 is unclear; there are no detections of either line but the upper limits are not very significant either. As discussed by Rigopoulou et al. (1997) in more detail, Mrk 231 is abnormal in the sense of having very small line to continuum ratios. Together with the warm IRAS 25/60μm colors these findings probably indicate that the 10-30μm emission is dominated by a small, optically thick warm dust source (see also the evidence on the lack of PAH emission below). The data for the weak sources IRAS 15250 and IRAS 19254 are poor so that the upper limits to [NeV] and [OIV] are not significant.

*We thus conclude from the high excitation line measurements that there is definitely one (Mrk 273) and probably a second (Mrk 231) ULIRG (out of 12) that are predominantly powered by a hard, AGN-like radiation field. The majority of the ULIRGs (7 out of 10, excluding the 2 ULIRGs without line detections), however, have line ratios that imply UV radiation fields softer than typical AGN radiation fields and are more similar to starburst galaxies*. In two sources (NGC 6240 and IRAS 23128) signs of an AGN are present in the ISO SWS spectra although star

formation appears to dominate the luminosity. Our new results thus strengthen the conclusions of our initial ISO-SWS spectroscopy of ULIRGs (Sturm et al.1996, Lutz et al. 1996a) and put them on a firmer statistical footing. We wish to emphasize again that *our data do not exclude that (most or even all) ULIRGs have central AGNs*. Our data do, however, provide strong evidence that (moderately extincted) AGNs are not the dominant sources of luminosity in 70 to 80% of the cases we have studied so far. In recent optical and near-infrared spectroscopic studies of ULIRGs Clements et al. (1996a,b) and Veilleux et al. (1995, 1997) likewise found that 25% to 35% of their sources contained AGNs.

## *3.4 Distinguishing between AGNs and Starbursts with PAH Features*

Unidentified emission bands at 3.3, 6.2, 7.7, 8.7 and 11.2μm (etc.) are probably emitted by large, carbon-rich ring molecules (e.g. polycyclic aromatic hydrocarbons [PAHs]) or very small, amorphous carbon dust grains that are exposed to moderately intense UV radiation (cf. Puget and Leger 1989, Duley and Williams 1991). While the actual carrier of these bands is still uncertain, we will, for simplicity, refer to them as 'PAH' features for their most popular identification. Observations in our Galaxy show that the PAH features originate in PDRs, at the interfaces between HII regions and dense molecular clouds (cf. Allamandola et al. 1995, Verstraete et al. 1996) but also in diffuse 'cirrus'/HI gas (Mattila et al. 1996). In a classical study of about 60 galaxies of different types Roche et al. (1991) have demonstrated empirically that PAH features are prominent in HII/starburst galaxies but are faint or absent in classical AGNs. With ISO the PAH features (especially the prominent 7.7/8.7 μm complex) can now be studied fairly easily also in the more distant ULIRGs.

Fig.4 is a plot of the strength of the 7.7μm PAH feature (see the footnotes in Table1 for a definition) for the 15 ULIRGs, as well as for classical AGNs starburst templates and Galactic HII regions/PDRs, including two positions in the Galactic center. Our data of the template galaxies are in excellent agreement with the general conclusions of Roche et al. (1991). The 7.7 μm feature is strong in dusty environments with moderately strong UV radiation fields, such as starburst galaxies and PDRs. The energy flux of the UV field in these regions ranges between 100 to a few $10^5$ times the energy flux of the radiation field in the solar neighborhood ($\chi_0 = 2 \times 10^{-4}$ erg s$^{-1}$ cm$^{-2}$ sr$^{-1}$). PAH emission is weaker (by factors of a few to more than an order of magnitude) in very strong radiation fields ($\chi \geq 10^6 \chi_0$). A good example is the central, dusty 'cavity' of the Galactic center (left-most open triangle in Fig.4). This can perhaps be understood as a result of destruction or transformation of the very small dust particles/large molecules in intense radiation fields (Puget and Leger 1989). As a group the AGNs in Fig.4 have about an order of magnitude weaker 7.7μm feature than the starburst templates. Composite galaxies (such as Circinus, NGC 7469, NGC 7582 and Cen A) range in between these extremes. ***Turning to the 15 ULIRGs in Fig. 4, all but Mrk 231, IRAS 23060 and IRAS 19254 have 7.7μm features that resemble in strength (and shape) those of the starburst templates.*** The measurements of the 7.7μm feature thus fully support the trend we have found from spectroscopy of the ionized gas that the majority of the ULIRGs (12 out of 15, or 80%) have the characteristics of galaxies powered predominantly by massive stars. The 'PAH strength' parameter is insensitive to extinction as long as the 7μm continuum is affected by the same dust obscuration as the emission features.

## *3.5 A New Diagnostic Diagram for ULIRGs: How Much AGN and How Much Starburst ?*

We now combine the information obtained from the high and low excitation lines (sampling the UV-irradiated, ionized gas) and from the strength of the PAH feature (sampling the UV-irradiated, hot dust) in a two-dimensional diagram (Fig.5 (left)). Here we plot the strength of the 7.7μm PAH feature on the horizontal axis and the 25.9μm [OIV]/ 12.8μm [NeII] (or 1.7 times 33.5μm [SIII]) ratio on the vertical axis. We have found that this diagnostic diagram is the best for empirically separating AGNs and galaxies powered by massive star formation. In this new diagram ULIRGs **on average** appear to be **composite objects** which, however, are located close to starbursts, especially if one takes into account that most of the [OIV]/[NeII] ratios are upper limits. From a simple 'mixing' model (AGN=([OIV]/[NeII]~1, PAH strength~0.04), starburst=([OIV]/[NeII]~0.02, PAH strength~3.6) we have attempted to assign more quantitatively the fractions of total luminosity contributed by AGN and starburst components. This is shown in Fig.5 (right) as a short-dash curve with specific AGN fractions indicated as crosses. In this model ULIRGs on average have a 5 to 30% AGN contribution with 70 to 95% coming from star formation. In Mrk 231 and Mrk 273 the AGN contribution may range between 40 and 80%.

Assuming now on the basis of the diagnostic diagram in Fig.5 that ULIRGs are on average dominated by stellar UV radiation from recently formed, massive stars, the next question is whether this assumption is consistent with the ratio of bolometric to Lyman-continuum luminosity in these galaxies.

### 3.6 $L_{bol}/L_{Lyc}$: Are ULIRGs Similar to Starburst Galaxies ?

Hydrogen ionization rates $Q_{Lyc}$ ($s^{-1}$) and Lyman continuum luminosities $L_{Lyc} = L(\geq 13.6 \text{ eV}) = Q_{Lyc}$ $<h\nu_{Lyc}>$ can be inferred directly from nebular hydrogen recombination lines. For this purpose we have taken case B recombination theory with an electron temperature of $T_e \sim 7500$ K and $n_e \sim 300$ $cm^{-3}$. We have adopted an average energy of a Lyman continuum photon of $<h\nu_{Lyc}> \sim 16$ eV (appropriate for galaxies powered by stars) and we have assumed that the ionized medium is ionization (and not density) bounded. We have assumed further that the infrared luminosity $L_{IR}$ of a galaxy (as derived from IRAS fluxes) is a reasonable measure of the bolometric luminosity $L_{bol}$ (excluding old stars) that has been absorbed and processed by interstellar dust and thus can be directly compared to $L_{Lyc}$. $L_{bol}/L_{Lyc}$ thus is a measure of the excitation and evolutionary state of the galaxy and its UV radiation field. We have first determined $L_{bol}/L_{Lyc}$ for ULIRGs using near-infrared recombination lines (e.g. 2.1μm Brγ) and near-infrared/optical screen extinctions. The results are shown in the top inset of Figure 6. Here we have compiled galaxies from the data of Goldader et al. (1995) with extinctions given by these authors. For comparison we show the corresponding $L_{bol}/L_{Lyc}$ histogram for the ISO starburst templates (middle inset) as taken from Table 2. A basically identical histogram emerges when the near-infrared data (and extinctions) of Goldader et al. (1995, 1997a,b) and Moorwood and Oliva (1988) are used for starburst galaxies. Starburst galaxies have a mean/median $<L_{bol}/L_{Lyc}> \sim 16$. The basic result of this comparison is ULIRGs have a much larger ratio of $L_{bol}$ to $L_{Lyc}$ than starburst galaxies ($<L_{bol}/L_{Lyc}>_{ULIRG} \sim 100$) when near-infrared/optical data are taken as the basis of estimating $L_{Lyc}$ (upper left). This apparent ***lack of ionizing luminosity in ULIRGs*** is well known and established in the literature (e.g. van der Werf et al. 1993, Goldader et al. 1995, Larkin et al. 1995, Armus et al. 1995). A common conclusion is that ULIRGs ***cannot be fit by normal starburst models where O stars dominate the $h\nu \geq 13.6$ eV radiation field*** [5].

---

[5] In the same context it is then argued that a large $L_{bol}/L_{Lyc}$ ratio better fits an AGN model. Superficially this is supported by the AGN data in Table 2 which result in an empirical $< L_{bol}/L_{Lyc}>_{AGN} \sim 30$ to 40 according to our definition. However, this definition is not suited for AGNs for which $<h\nu_{Lyc}> >> 13.6$ eV and the assumption of an

Next we estimate $L_{Lyc}$ in ULIRGs from our new ISO data. Columns 10 to 12 of Table 2 list our estimates of the ratios of far-infrared to Lyman continuum luminosities of the ULIRGs (and the other galaxies) we have observed with ISO, either derived directly from Brβ and/or Brα, or indirectly from the [NeII] 12.8μm and/or [SIII] 33.5μm luminosities and empirical conversions to $L_{Lyc}$. The latter we determined empirically from the average ratio of $L_{Lyc}/L([NeII])$ and $L_{Lyc}/L([SIII])$ in starburst galaxies (64 and 90 from Table 2). We have commented earlier that for galaxies dominated by typical stellar UV fields such a simple linear proportionality is expected from photoionization models. For solar abundances (Ne/H=1.1x10$^{-4}$ ) and log U=-1.5 to -2.5 the value of $L_{Lyc}/L([NeII])$ predicted by photoioinization models with the new NLTE, line blanketed models of Pauldrach et al. (1997) for $T_{eff}$~33-42 kK and $T_e$~7500 K is 100±15. This is in reasonable agreement with the empirical value but probably indicates a ~50% overabundance of Ne relative to solar. For $L_{Lyc}/L([SIII])$ the value predicted from theory with solar abundance (S/H=1.6x10$^{-5}$) is 36 with a small dispersion, 2.5 times lower than the empirical value. This indicates that in the galaxies we have observed the gas phase abundance of sulfur is about 0.4 times solar, in good agreement with infrared spectroscopy of Galactic HII regions (Simpson et al. 1995). Column 13 lists the ratio of L(FIR)~$L_{bol}$ to 75 times L([SiII]). The [SiII] line is a measure of the 8-13.6 eV luminosity which is comparable to $L_{Lyc}$ for hot stars and in fact a conversion factor of $L_{Lyc}/L([SiII])$~75 is again the average for the starburst galaxies in Table 2. Within the uncertainties ( ±50% to factor of 2) resulting from calibration, extinction correction and conversion to $L_{Lyc}$, the different estimators of $L_{bol}/L_{Lyc}$ agree reasonably well in most galaxies, including that using the [SiII] line. An exception to this good agreement is Arp 220 (see footnote 8). Taking, wherever possible, averages of all the HII region tracers

---

ionization bounded, unity coverage factor, case B HII region is usually not justified, or the extinction of the ionizing luminosity is underestimated. In reality <$L_{bol}/L_{Lyc}$>$_{AGN}$~1 (e.g. Netzer 1990).

(recombination lines, [NeII], [SIII],[SiII]) we list the resulting mean $L_{bol}/L_{Lyc}$ in column 14 of Table 2.

We plot the ISO derived values of $L_{bol}/L_{Lyc}$ for ULIRGs in the bottom inset of Fig.6. From the ISO data we infer on average a 3 times larger Lyman continuum luminosity for the ULIRGs than from the near-infrared data, resulting in a corresponding shift of the (Fig.6 (top)) ULIRG histogram toward the left. This difference is the direct result of the much larger extinctions implied by the mid-/far-infrared spectroscopy. T*he extinction corrected, Lyman continuum luminosities in ULIRGs derived from mid-IR recombination lines or low excitation fine structure lines are thus large enough to account for the total bolometric luminosities with UV radiation fields that are similar to starburst galaxies with recently formed massive stars dominating the 13.6 to 35 eV energy range.*

The last conclusion would not be expected if all ULIRGs were heavily extincted AGNs (see section 3.7). In that case both high and low excitation lines would be equally suppressed by the large extinction, as perhaps is indeed the case in Mrk 231. However, $L_{bol}/L_{Lyc}$ cannot be used as a quantitative diagnostic tool for distinguishing starburst galaxies and AGNs as their respective mean values of $L_{bol}/L_{Lyc}$ (or rather $L_{bol}/Q_{Lyc}$ where $Q_{Lyc} = L_{Lyc}/<h\nu_{Lyc}>$ ) are so close to each other (Table 2 and Fig.6).

If the main reason for the weakness of the near-infrared (and optical) hydrogen recombination line emission of ULIRGs is that intrinsically bright star formation regions are heavily obscured by dust but otherwise are similar to those found in starburst galaxies, one would expect that extinction free near-infrared indicators, such as the ratio of Brγ flux to K-band stellar flux (the equivalent width of Brγ), should be similar in ULIRGs and starburst galaxies. We test this hypothesis, again using the data bases of Goldader et al. (1995, 1997a,b), Moorwood and Oliva (1988) and this paper, and show the results in Fig.7. ULIRGs as a group range between pure starburst galaxies and AGNs. Several of the ULIRGs with $L_K/L_{Lyc} \geq 0.4$ have low CO indices, indicative of a significant non-stellar nuclear (AGN) K-band component (IRAS 05189-2524, Mrk 231, IRAS 23365+3604: Goldader et al. 1995, UGC 5101: this paper). If the observed K-band fluxes are corrected for this non-stellar emission, these objects move to the left. In the case of UGC 5101 (observed $L_K/L_{Lyc} = 0.4$ ) the correction is about one bin ($L_K$(stellar)/$L_{Lyc}$ ~0.2). Considering this effect Fig.7 is indeed in good agreement with the hypothesis that ULIRGs are more similar to starburst galaxies than AGNs. The one obvious exception with the highest $L_K/L_{Lyc}$ value among the ULIRGs is NGC 6240. There the large K-band flux is of stellar origin. The NGC 6240 merger system is known to have an unusually high stellar velocity dispersion (σ~350 km/s, Lester and Gaffney 1994, Doyon et al. 1994b, Tecza et al. 1997) indicating a particularly high central stellar/gas concentration and signaling that the system is in a late, post-starburst phase (van der Werf et al. 1993, Moorwood 1996, Tecza et al. 1997).

### 3.7 Can AGNs Be Hidden even at 30μm?

The only way out of our main conclusion that the majority of the ULIRGs in Table 2 is powered predominantly by star formation is to postulate that ULIRGs have dominating AGNs that are all hidden even at 10 to 35μm. At these wavelengths $\tau_{dust}$ ~1 corresponds to (screen) extinctions of $A_V$~ 50 to 100 magnitudes (N(H+2H$_2$)~1 to 2x10$^{23}$ cm$^{-2}$). To hide completely a powerful 10-35μm source requires $\tau_{dust}$~ 2 to 3, or N(H+2H$_2$)~2 to 6x10$^{23}$ cm$^{-2}$, spread over the size scale of the mid-/far-IR continuum source. Such large dust extinctions are certainly conceivable on the basis of the H$_2$/dust column densities inferred from millimeter CO J=1-0 line emission and submillimeter dust emission in the ULIRGs we have observed ($A_V$(CO, submm)~$10^2$ to $10^3$, Rigopoulou et al. 1996, Solomon et al. 1997, Scoville, Yun and Bryant 1997, Downes and Solomon 1997). The 'hidden AGN' explanation, however, is not very likely for several reasons. First, as discussed already in the last section, the ionizing luminosities we deduce from the ISO low-excitation lines using the (moderately large) extinctions deduced from the mid-infrared data are fully sufficient for explaining the far-infrared continuum as being reprocessed stellar luminosity. Taking the average ratio of $L_{bol}/L_{Lyc}$ ~16 in our starburst templates as characteristic of star formation regions in ULIRGs as well, we can infer that *in about half of the ULIRGs* in Table 2 *at least half of the luminosity must come from star forming regions*. This does not require any ageing of the starbursts beyond what is seen in (the generally smaller) starburst templates or a correction for absorption of Lyman continuum photons by dust. If we allow for additional ageing and/or dust absorption, both the relative number of ULIRGs and the fraction of their luminosity powered by star formation would increase (see section 5 below).

Second, 1 to 10 keV X-ray observations of ULIRGs do not reveal strong emission with a Seyfert-like $L_X/L_{IR}$ ratio (Rieke 1988, Mitsuda 1995, Rigopoulou et al. 1996, Kii et al. 1997). 5-10 keV X-rays can penetrate to N(H)~$10^{24}$ cm$^{-2}$. In the case of Arp 220, recent ASCA observations indicate a ratio of 2 -10 keV luminosity over 8-1000μm luminosity of $\leq 10^{-3.7}$ (Mitsuda 1995, Kii et al. 1997). For comparison, Seyfert 1 and 2 galaxies have $< L_X/L_{IR} > ~ 10^{-1}$ and $10^{-2}$, respectively (Boller et al. 1997, Awaki et al. 1991). ASCA observations *do* provide evidence for powerful AGNs in some ULIRGs. NGC 6240 has a highly absorbed hard X-ray source with $L_X/L_{IR}$ ~$10^{-2}$ to $10^{-1}$, indicating the presence of substantial AGN activity (Mitsuda 1995, Kii et al. 1997). Our ISO detection of [OIV] line emission in NGC 6240 probably provides evidence for an AGN as well (although a contribution of shocks to the [OIV] emission in this advanced merger needs to be investigated). IRAS 23060+0505 is associated with a hard X-ray source with $L_X/L_{IR}$~$10^{-2}$ (Brandt et al. 1997). Our ISO data also identify this source as an AGN candidate on the basis of the 'PAH strength' criterion. In the near-infrared IRAS 23060+0505 shows an absorbed/scattered Seyfert 1 nucleus (broad line emission, Veilleux et al. 1997, Young et al. 1996) with an extinction comparable to that estimated from the X-ray data, $A_V \sim$ a few tens.

Third, we show below that the one definite AGN we have found on the basis of high to low excitation line ratios (Mrk 273) is in no way unusual in terms of (low) extinction, X-ray luminosity or optical/IR/radio morphology. Finally, only a small subset of ULIRGs shows strong mid-IR continuum emission ('warm' ULIRGs: 25/60μm flux density ratio $\geq 0.2$, Sanders et al.1988a). If all ULIRGs were dominated by deeply embedded AGNs ($N(H+2H_2) \geq 10^{24}$ cm$^{-2}$) the most likely configuration of the obscuring medium according to standard 'unified' schemes would be a very dense, compact (1 to 100 pc) circum-nuclear molecular torus or warped disk (e.g. Antonucci 1993). This circum-nuclear torus would convert most of the short-wavelength AGN luminosity into a strong, fairly featureless 5 to 30μm dust continuum (excepting 10/20 μm silicate absorption, see model calculations by Rowan-Robinson and Crawford 1989, Rowan-Robinson and Efstathiou 1993, Pier and Krolik 1993, Granato and Danese 1994). This is indeed seen in the subset of 'warm' ULIRGs (e.g Mrk 231) which also often show clear evidence for AGN activity at optical or near-IR wavelengths (Veilleux et al. 1997).[6] To hide the dust tori even at 30μm would, as a rule, require a very special (edge on) geometry (Pier and Krolik 1993), or yet greater dust column densities spread evenly over a larger (~kpc), more or less spherical volume. However, fairly basic considerations of the scale height of a molecular disk under its own gravity, combined with the observed thermal and kinematic properties of the molecular gas, indicate that the molecular gas disk in Arp 220, for instance, is fairly thin (H~20pc, Scoville, Yun and Bryant 1997). That inferred thickness is smaller than the minimum 'blackbody' radius $R_{bb}$ that would be required to hide a point-like nucleus in Arp 220 even at $\lambda \sim 30\mu m$,

$$R_{bb} = 80 \quad (L/1.3 \times 10^{12} \, L_o)^{1/2} \quad (100 \, K/T_{dust})^2 \quad (pc) \quad\quad\quad (1).$$

---

[6] Such a model of a deeply hidden, compact mid-IR source has also been proposed for Arp 220 (and IRAS 08572+3915) by Dudley and Wynn-Williams (1997) on the basis of a deep 9.7μm silicate absorption deduced from ~5.5" aperture, ground based spectroscopy they and Smith, Aitken and Roche (1989) have carried out. The Smith et al. data and our Arp 220 ISOPHOT-S spectrum are in excellent agreement, however, both in terms of shape and absolute flux (after correction for the 40% chopping loss mentioned in 2.1). The difference is that the ground based data only cover the 7.8μm to 13μm atmospheric window. The data in Fig.2 definitely show that the sharp decrease between 7.8 and 9μm is mainly due to PAH emission and not due to silicate absorption. From our data we estimate a silicate optical depth of ~1.5 (instead of 5 to 9 in Dudley and Wynn-Williams and Smith et al.), indicating that a deeply absorbed, 'protostellar' model is not required or justified for Arp 220.

Finally the nucleus (or nuclei) in Arp 220 and UGC 5101 (see section 4 below) and other ULIRGs are actually seen in the near-IR (e.g. Graham et al. 1990, Majewski et al. 1993, this paper) indicating that the $\leq 10^2$ pc circum-nuclear regions are not fully covered by the thick ($N(H+2H_2) > 3$ to $10 \times 10^{24}$ $cm^{-2}$) molecular gas layer (Scoville, Yun and Bryant 1997). ***It is therefore not likely that all ULIRGs are powered by heavily obscured AGNs.***

### 3.8 Could the Relative Importance of AGNs and Starbursts in ULIRGs Be Luminosity Dependent ?

Based on optical spectroscopy Veilleux et al.(1995), Kim et al.(1995) and Kim (1995, as referred to in Figure 5 of Sanders and Mirabel 1996 and in Veilleux et al. 1997) have found a rapid increase in the fraction of Seyfert nuclei among ULIRGs of luminosity greater than $10^{12.3}$ $L_o$ and with increasing S(25)/S(60) ratio. The question thus arises whether the relative importance of AGNs and starbursts in ULIRGs could be a strong function of luminosity. Our present sample unfortunately cannot provide a conclusive answer. Table 2 contains four objects with $L \geq 10^{12.3}$ $L_o$. According to the PAH and/or [OIV]/[NeII] criteria two of them are in the AGN category (Mrk 231, IRAS 23060), two are probably in the starburst category (IRAS 17208, IRAS 20100), implying a 50:50 distribution. This may be in fact consistent with the findings of Veilleux, Kim and coworkers but is statistically obviously too uncertain to be a fair test (pro or con). We are presently carrying out further ISOPHOT-S observations to test such a luminosity dependence.

# 4. CASE STUDIES: UGC 5101 AND ARP 220

In the last section we have explored the global characteristics of ULIRGs with ISO aperture spectroscopy. Our central finding - that star formation dominates the luminosity of most ULIRGs - leads to the obvious prediction that extended, powerful star forming regions should be resolvable on scales $\leq$ a few kpc (typically a few arcseconds at the distances of the ULIRGs). For this reason we discuss in this section high resolution near-infrared and mm imaging spectroscopy we have carried out of one galaxy in the sample (UGC 5101) and we summarize similar observations of the prototypical ULIRG, Arp 220. We will show in this section that the observed small scale characteristics are in excellent agreement with the predictions from the ISO survey.

UGC 5101 is one of the ten ultra-luminous galaxies identified in the IRAS Bright Galaxy Survey (BGS: Soifer et al. 1986). In many respects it is a very typical representative of its class. It has a 8-1000μm luminosity of ~$10^{12}$ $L_o$ with a fairly typical 'cool' far-infrared color temperature (~49 K, Solomon et al. 1997). It contains a large mass (~$2x10^{10}$ $M_o$) of molecular hydrogen (Solomon et al. 1997). Its optical spectrum puts it in the Seyfert 1.5 class (Sanders et al. 1988a, Veilleux et al. 1995). It has a disturbed optical morphological shape, with a pronounced tail ~40" west of the single nucleus, as well as a ring at a radius of ~20" (Sanders et al. 1988a). Lonsdale et al.(1993) found that 20% of UGC 5101's 1.6 GHz radio flux is in a $\leq$0.2" compact nuclear source with a brightness temperature of at least $10^7$ K. At VLBA resolution this emission breaks up into several bright knots that are just resolved at ~6 milli-arcsec resolution (Lonsdale, Lonsdale and Smith 1995). Apart from Mrk 231, UGC5101 has the strongest nuclear radio source in the bright source ULIRG sample observed by Lonsdale et al.(1993).

In the following section, we will discuss the results of our H- and K-band imaging spectroscopy of UGC 5101 obtained with 3D and of the IRAM interferometry of 2.6 mm CO 1-0 line emission in that source.

## 4.1 Near-Infrared Maps and Spectra

Figs.8, 9 and 10 show line and continuum maps constructed from the 3D data cubes and Fig.11 shows H- and K-band spectra of selected positions. The basic results are as follows.

The K-band **continuum** has a core-halo structure with an unresolved central source (observed FWHM 0.94", implying an intrinsic size $\leq$0.5") superposed on an elliptically shaped source of FWHM 2.5"x2.0" with major axis at position angle $79^0$ east of north (Fig.8). Within the 12.8" 3D field the two components contribute about equally to the total K-band flux density of 23 mJy, but on larger scales the extended K-band component is only the brightest part of a stellar disk that is surrounded by a ring/jet structure on a scale of $\geq$30" (Sanders et al. 1988a). Extended and compact sources taken together are fairly red (H-K=0.97 or $F_\lambda$~const($\lambda$), $S_\nu$~$\nu^{-2}$) but, in addition, 20% of the K-band flux of the nuclear source in the central pixel comes from a very red ($F_\lambda$~$\lambda^2$, $S_\nu$~$\nu^{-4}$) ) component that can be best traced just below the long wavelength edge of the K-band . Since the total L'-K color of the source is also red (K-L'=1.34, Sanders et al. 1988a) it is very likely that the 3.7$\mu$m excess is caused by this red nuclear component. In analogy with Seyfert galaxies with similarly red nuclear near-/mid-infrared sources (NGC 1068, NGC 7469, Circinus) we interpret this excess to be due to hot dust (500-1000 K) in the immediate vicinity of an active nucleus in UGC 5101 (see Thatte et al. 1997, Genzel et al. 1995 and Maiolino et al. 1997 for recent data on the hot dust sources in NGC 1068, NGC 7469 and Circinus). The size of the hot dust source in NGC 1068 and Circinus is less than two to three parsecs (Thatte et al. 1997, Maiolino et al. 1997). This is comparable to the size of the central, flat spectrum radio core S1 in NGC 1068 which may be coincident with the inner rim of the putative circum-nuclear torus or warped disk (Gallimore et al. 1997).

## *4.2 Hot Molecular Hydrogen and Ionized Gas Emission*

Vibrationally excited molecular hydrogen line emission ($H_2$ v=1-0 S(1), S(2) and S(3)) and HI P$\alpha$ and Br$\gamma$ line emission (Fig.8) come from a somewhat elongated source (FWHM 1.5"x1.25" also along p.a.79°) of total size ~2.5". Stepping through individual velocity channels it is apparent that the neutral and ionized gas originate in a rapidly rotating ring (or disk) surrounding the nucleus (Fig.8). The projected rotation velocity is ≥150 km/s and the ring's radius is about 0.45" (350 pc). There does not appear to be a strong compact P$\alpha$ (or $H_2$ S(1)/S(3)) source associated with the nucleus itself. The P$\alpha$ and $H_2$ emission seen directly toward the nuclear position is near the rest frame velocity (Fig.8). This is exactly what is expected for the projection of a rotating gas ring. The ratio of P$\alpha$ to Br$\gamma$ has a maximum at the nucleus (~15±3) but decreases to values of ~8±3 in the ring on either side of the nucleus. This implies relatively low extinction toward the nucleus (the measurement is consistent with $A_V$=0 but the error bars allow values as high as $A_V$=7 in a screen model) while the extinction toward the ring appears to be higher ($A_V$~20±10 for a screen model and higher if gas and dust are mixed). The relative strengths of the observed $H_2$ lines, v=1-0 S(1), S(2) and S(3), are in good agreement with that of thermally excited, dense (n($H_2$)≥$10^4$ $cm^{-3}$) gas of temperature ≥1000 K. The integrated P$\alpha$ flux over the central 5" is 6.3x$10^{-21}$ W $cm^{-2}$, as calibrated from the K-band continuum flux density of 22 mJy (K=11.13) of Sanders et al. (1988a).

## *4.3 Stellar Absorption Features in the H- and K-Band*

The 2.3μm CO v=0-2 and 1.6μm CO v=3-6 absorption flux maps in Figs.8 and 9 also show the same ring/disk morphology and kinematics as the interstellar gas, with no or little evidence for a concentration toward the nucleus. There are peaks in the CO 0-2 absorption flux (proportional to the number of late type stars if the stellar type is constant) 0.35" east and (north-) west of the nucleus (Fig.8). The (north-) western peak is more prominent at CO 0-2 and dominant at CO 3-6, suggesting an extinction gradient, with the eastern side of the galaxy more obscured than the western side. In the K-band the 0-2 CO band-head at 2.294μm (rest wavelength) is the most prominent feature but NaI, CaI and FeI absorption features can also be (marginally) recognized (Fig.11). In the H-band the CO v=0-3, 1-4, 2-5, 3-6, 4-7 and 5-8 band heads are seen as well as absorption due to 1.59 μm SiI (Fig.11). These absorption features are characteristic for late type K/M giants or supergiants. The equivalent width of the 2.29μm CO band, as estimated from integrating the absorption over about 900-1200 km/s centered near the absorption peak, has a maximum of 13±3Å in the ring and decreases to less than than 3Å toward the nucleus. This decrease is caused to a large extent (but not exclusively) by dilution due to the nuclear hot dust emission. The CO and SiI absorption features in the H-band show a similar behavior. The CO 3-6 and 1.59μm SiI equivalent widths in the ring have peak values of 7±1 and 5-6 Å, respectively (again integrating over about 900-1200 km/s). The nuclear values are about half that, and the flux maps show that the decrease is not just caused by increased non-stellar dilution. Combining the equivalent widths of the H and K-band CO and SiI features and comparing to the stellar template observations of Origlia, Moorwood and Oliva (1993) and Oliva et al.(1995) we find that the stellar population is characterized by an effective temperature of 3700±400 K (~K4/5).

The circum-nuclear disk/ring is also the source of spatially extended, non-thermal radio continuum emission with a fairly steep spectral index (Condon et al. 1991b, Sopp and Alexander 1991). The 0.25" 8.4 GHz radio continuum map of Condon et al. (1991b) also shows the core-halo structure of the near-infrared tracers and the extended radio emission traces the circum-nuclear ring very well (Fig.10). After correction for the central compact radio source (<0.2", Lonsdale, Smith and Lonsdale 1993), the ring's radio emission of UGC 5101 falls exactly on the far-infrared-radio continuum relationship for normal and starburst galaxies (Condon et al. 1991a, Helou et al. 1985). This relationship is usually interpreted in terms of the global proportionality of (massive) star formation and supernova rate which in turn correlates with radio continuum emission from supernova remnants.

Taken together, the intense HI recombination line emission, the $H_2$ emission, the extended non-thermal radio continuum emission and the fairly deep stellar absorption features from late type stars strongly suggest that *the circum-nuclear ring is the location of a powerful starburst*.

### 4.4 CO 1-0 Millimeter Interferometry

Fig.12 shows the spectra (A), spatial maps (B,C) and a major axis, spatial-velocity map (D) of the 2.6mm CO J=1-0 line emission toward UGC 5101, as obtained with the IRAM Plateau de Bure mm-interferometer at 2.1"x1.6" (FWHM) resolution. The integrated molecular line emission comes from a compact source that is barely resolved at half power level but with fainter emission extending to ±5" from the nucleus along the same position angle as the near-infrared stellar emission. Like the stars and the near-infrared tracers of the gas, the molecular emission clearly comes from an inclined rotating gas disk. This is apparent from the velocity profiles (A), the channel maps of peak blue- and red-shifted gas (B), and from the spatial-velocity map (D) displayed in Fig.12. Emission maxima in the blue-/red-shifted gas at ±0.75" from the nuclear position mark the peak radius of the circum-nuclear disk/ring. The minor to major axis ratio in the well resolved lower intensity isophotes of the integrated CO line emission ( C) is 0.57±0.03. Taking into account the effects of beam smearing, we infer from this ratio an inclination of about $60^o$.

Owing to the excellent spectral resolution of the CO mm-data (20 km/s) the position-velocity diagram in Fig.12 (D) can be used to accurately determine the kinematic properties of the rotating disk. Employing the same rotating disk simulation program as described in Tacconi et al. (1994), we have modelled the position-velocity map in Fig.12 with an inclination $60^o$, axisymmetric rotating disk/ring. We find a good fit to the data for the combination of a bright ring centered at R=0.55" (420 pc) with a Gaussian FWHM radial width of 0.5", plus a 30 times lower surface density disk of Gaussian FWHM radial scale length 3" (2.2 kpc). With this radial distribution of surface density the position-velocity map is well fitted with a rotation curve of

$$v_{Rot}( R )= 322 ~( R/0.5")^{-0.06} ~(\sin(60^o)/\sin(i)~) \qquad (km/s) \qquad (2).$$

The local (turbulent) velocity width is about 80 km/s FWHM. This rotation curve implies a dynamical mass of $9.2 \times 10^9 ~(R/0.5")^{0.88}~(M_o)$. Within the outer radius of the bright CO/starburst ring at R~2"(1.5 kpc) the dynamical mass thus is $3.1 \times 10^{10} ~M_o$. Downes et al.(1993) and Solomon et al. (1997) have argued that the far-infrared and CO 1-0 emission of ULIRGs are both optically thick. In this case the observed far-infrared spectral energy distribution of UGC5101 ($T_{color}$ ~ 49 K, $L=10^{12} L_o$) implies an equivalent blackbody radius of $R_{bb}$~ 0.4", quite comparable to the observed radius of the starburst ring. $R_{bb}$ is a lower limit to the true mean radius of far-infrared emission. It is thus plausible from this argument as well that a major fraction of UGC 5101's far-infrared luminosity emerges from a region of size comparable to (or larger than) that of the starburst ring. The characteristics of the circum-nuclear disk/ring in UGC 5101 are similar to those found in other ULIRGs by Downes and Solomon (1997) and Scoville, Yun and Bryant (1997).

The interferometer data of UGC 5101 can be used to show that the dense interstellar gas accounts for about 50% of the entire dynamical mass in the central few kpc and that the CO 1-0 intensity to $H_2$ gas mass conversion factor is within a factor of 2 of the value derived in the Galaxy. Details can be found in Appendix A.

With a molecular mass in the brightest part of the circum-nuclear disk/ring of $10^{10.0\pm0.3}$ $M_o$ within $R\le 1$ kpc (Appendix A, Table 4) the average $H_2$ surface density corresponds to $N(H_2) = 10^{23.3\pm0.3}$ $cm^{-2}$ or $A_V$ (CO)~200. The average 'mixed case' extinction to the star forming complexes in UGC 5101 is $A_V$ (IR, mixed)~50 (Table 2) while the equivalent 'screen' extinction would be $A_V$(IR, screen)~5 to 7. Hence $A_V$(CO)$\ge$$A_V$(IR, mixed)$>>$$A_V$(IR, screen)[7] indicating that the circum-nuclear starburst in UGC 5101 - as in nearby starburst galaxies - is highly clumped with dense molecular clouds and (super-) clusters of young stars being intimately mixed.

## *4.5 Summary of observations of Arp 220*

We have already referred several times to various observations of the closest and best studied ULIRG, Arp 220 (distance 73 Mpc, 1"=350 pc). Here we want to briefly summarize a few key high resolution observations which give insight on the processes happening on scales less than a few kpc. These observations support the notion that also Arp 220 is powered mainly by star formation.

---

[7] Likewise the equivalent mixed case extinction for Arp 220 is $A_V$(IR, mixed)~1000 or $N(H_2)$~$10^{24}$ $cm^{-2}$, again comparable to the column density of molecular hydrogen estimated from mm CO observations.

- Graham et al. (1990) have shown that the near-infrared emission shows two prominent maxima, separated approximately 320 pc in east-west direction. Recently a beautiful new image has been taken with the NICMOS camera onboard HST (Scoville et al. 1998). Graham et al. interpret these two peaks as the two nuclei of the merging galaxy pair. The nuclei are highly extincted, as they are not visible on Hubble Space Telescope V-, R- and I-band images, and the recent multi-color, near-infrared NICMOS image shows strong color gradients (Scoville et al. 1998).

- Armus et al. (1995) have presented low spectral resolution, K-band spectra of both nuclei taken with the Keck telescope. They find that young red supergiants contribute significantly (~50%) to the near-infrared emission of both nuclei and there is no evidence for hot dust emission from a compact nuclear region(s). The equivalent width of Brγ is consistent with a model of recent (~$10^7$ years) massive star formation. However, the Lyman continuum luminosity derived from the Brγ flux, corrected for a visual extinction of ~10 magnitudes, as obtained from the Brγ/Pfβ line ratio (Larkin et al. 1995) in a screen model, fails the Lyman continuum luminosity of a ~$10^7$ year starburst model with $L_{Bol}$~$10^{12.1}$ $L_o$ by about 2.8 magnitudes. Armus et al. therefore conclude that the power source of Arp 220 ist still hidden at 2μm. This problem is resolved when the new ISO-derived extinction is used instead ($A_V$~45, Sturm et al. 1996, this paper). The difference between the K-band extinction used by Armus et al. and that implied by our ISO SWS data is 3.4 magnitudes, more than enough to fully account for the far-infrared luminosity with a starburst model.

- Smith et al. (1998) have presented 3x8 milli-arcsec resolution, 18cm VLBI observations of the two radio nuclei of Arp 220. They find that ~3% of the total radio flux of the Arp 220 system comes from more than a dozen, unresolved point sources, spread over the two nuclei, the rest of the flux being in lower surface brightness, extended emission. Smith et al. interpret this population of bright nonthermal, compact sources as very luminous radio supernovae ('hypernovae') that have recently gone off in a dense medium. This interpretation is consistent with the expected supernova rate in a simple starburst model of Arp 220.

- The two nuclei are also prominent in molecular gas but appear to be embedded in a larger scale disk/ring system (Scoville et al. 1997, Downes and Solomon 1997).

- There is no evidence for an obscured hard X-ray source of $L(2\text{-}10 \text{ keV})/L_{FIR} \geq 10^{-3.7}$ (Mitsuda 1995, Kii et al. 1997).

- The LINER spectral characteristics of Arp 220 in the visible band may be related to the (starburst driven) superwind associated with Arp 220 (Heckman, Armus and Miley 1990).

# 5. EVOLUTIONARY STATE OF CIRCUM-NUCLEAR STAR FORMATION IN ULIRGS

In this section we explore in more detail the characteristics and evolutionary state of star formation regions in ULIRGs. Table 3 summarizes the basic properties of the circum-nuclear starburst in UGC 5101.

For this purpose we use the far-infrared luminosity $L_{IR}$, the ratio $L_{IR}/L_{Lyc}$ and the (extinction free) ratio of (stellar) K-band luminosity ($L_K$ is defined to be the luminosity in the K-band (1.9 to 2.5µm) to $L_{Lyc}$ as the main global parameters to characterize the star formation. For comparison we also use, as an additional global parameter, the ratio of supernova rate ($\nu_{SNR}$) to infrared luminosity. We estimate the supernova rate $\nu_{SNR}$ (yr$^{-1}$) from the (non-thermal) radio flux, as deduced by Huang et al. (1994) from observations of M82,

$$\nu_{SNR} \, (yr^{-1}) \; = \; 3.4x10^{-6} \, (\nu_{5GHz})^{\alpha} \, S(mJy) \, D_{Mpc}^{2} \qquad (3).$$

Here S(mJy) is the (non-thermal) radio flux density at the radio frequency in units of 5 GHz, $\nu_{5GHz}$, for a galaxy at distance D in Mpc. As described in Lutz et al. (1996a), we compare the observed global parameters to those predicted by a model based on the star cluster evolution code STARS (Krabbe, Sternberg and Genzel 1994, Genzel et al. 1995, Kovo and Sternberg 1997). STARS calculates the number of stars in the various parts of the Hertzsprung-Russell diagram ($T_{eff}$, L) as a function of time for different star formation histories and then determines the global parameters of that star cluster. The parameters characterizing the star formation history we express in terms of the decay time $\Delta t$ of an exponential model: $R_*(t) = R_o \exp(-t/\Delta t)$. A very sharp ($\delta$-) burst has $\Delta t \ll 10^7$ years, and a constant star formation has $\Delta t \gg 10^7$ years. We assume a Salpeter initial stellar mass function where the number of stars formed between m and m+dm is given as $dN(m) \sim m^{-\gamma} dm$, with $\gamma = 2.35$ between $m_l = 1\ M_o$ and $m_u$. We consider $m_u \sim 50$ ( to 100 ) $M_o$ as the most likely value of the most massive stars that appear to be produced in starburst templates (Doyon et al. 1994a, Rieke et al. 1993, Moorwood and Oliva 1994, Krabbe et al. 1995, Genzel et al. 1995, Lancon and Rocca-Volmerange 1996, Förster-Schreiber et al. 1997, Böker et al. 1997). The model then calculates the integrated properties ($L_{bol}$, $L_K$, $L_{Lyc}$, $\nu_{SNR}$, $M_*$ etc.) based on the distribution of stars in the HR diagram as calculated by the stellar evolution models of Meynet et al. (1994) and Schaerer et al.(1993) for solar metallicity and normal mass loss. Fig. 13 shows the resulting model parameters for $m_u = 50\ M_o$ as a function of t and $\Delta t$. We overlay on these plots the locations for UGC 5101 and other ULIRGs , as well as the average location of the starburst templates.

The observed integrated properties of the ISO starburst galaxies imply burst ages varying in different sources from values under 10 million years ('interaction zone' in NGC 4038/39, Kunze et al. 1996, Vigroux et al. 1996, Fischer et al. 1996) up to 100 million years (Fig.13). The median is about t=7-10 million years for $m_u=50\ M_o$ and t=20 million years for $m_u=100\ M_o$. The burst duration $\Delta t$ is not well constrained but may be comparable to the burst ages and may range between 10 and 20 million years. UGC 5101 fits the trend of the starburst templates and appears to be an active, moderate age starburst. Good fits for UGC 5101 can be obtained for $m_u \sim 50\ M_o$, t$\sim \Delta t \sim 10$ to 30 million years (Fig.13). The other ULIRGs (including Arp 220)[8] are similar, with a significant implied spread of ages (10 to 100 million years).

---

[8] For Arp 220 we estimate $L_{Lyc}$ ranging from $1.3 \times 10^{10} L_o$ ([SIII] and [SiII]) to $3 \times 10^{10} L_o$ (from [NeII]), to $8.7 \times 10^{10} L_o$ (from Br$\beta$, Br$\alpha$, Pf$\alpha$) (Table 2). This large range of the different tracers is worrisome. From the mm/submm-spectrum of Arp 220 Scoville, Yun and Bryant(1997) and Downes and Solomon (1997) estimate a 110 GHz thermal flux density of 10 mJy and

One major effect not taken into account in our analysis so far would tend to push the $L_{bol}/L_{Lyc}$ values of ULIRGs (and starbursts) to lower values. This is the absorption of Lyman continuum photons by dust inside HII regions. This effect is well known from far-infrared and radio continuum observations of compact HII regions in our own Galaxy, where the observed 'infrared excess' may suggest that half (or more) of the Lyman continuum luminosity is directly absorbed by dust (e.g. Furniss, Jennings and Moorwood 1974, Genzel et al. 1982, Wood and Churchwell 1989). Correction for this effect in the dusty environment of ULIRGs could decrease the inferred $L_{bol}/L_{Lyc}$ by perhaps as much as a factor of 2 over the observed values (see discussion in Moorwood 1996). In that case starburst ages would drop by a factor 1.5 to 2. For the starburst galaxies, the supernova rates estimated from the radio flux and the Huang et al. (1994) conversion factor are compatible with the model predictions. However, the ULIRG radio fluxes (e.g. UGC 5101) indicate supernova rates that appear to be a factor of ~2 too high unless one chooses $\delta$-burst models tuned to just the right age (~15 to 30 million years). This likely discrepancy could be easily removed by adjusting the empirical conversion factor in equation (3) downward by that factor. The physical reason for this discrepancy may be that the interstellar medium in which supernovae explode is much denser than in M82 and that unusually bright radio supernovae ('hypernovae') contribute significantly, as in Arp 220 (Smith et al. 1997). In summary then the global parameters $L_{Bol}/L_{Lyc}$, $L_K/L_{Lyc}$ and $\nu_{SNR}/L_{bol}$ in ULIRGs *fit models of evolving starbursts pretty well but probably indicate somewhat older burst ages than in lower luminosity, nearby starburst galaxies*. Our models favor time restricted bursts of age and duration of less than a few tens of millions of years.

---

13 mJy, respectively. This corresponds to $L_{Lyc}=6-7.8 \times 10^{10}(T_e/7500 K)^{-0.45}$. From observations of the H92$\alpha$ recombination line Zhao et al.(1996) obtain $L_{Lyc}$~4 $\times10^{10}$ $L_o$. The radio observations thus range in the middle of the ISO range and favor more the [NeII]/recombination line estimates. We take the median of these different values as the best estimate of the Lyman continuum luminosity for Arp 220: $L_{Lyc}=3.5 \times10^{10}$ $L_o$, or $L_{bol}/L_{Lyc}$ ~ 37 (Table 2)

# 6. CONSEQUENCES FOR THE EVOLUTION OF ULTRA-LUMINOUS IR GALAXIES

While our new data strongly support the 'starburst'- scenario for most of the ULIRGs observed so far, they show also that at the same time AGNs are important. The ISO observations of Mrk 231 will be discussed in more detail in a forthcoming paper (Rigopoulou et al. 1997). It is thus perhaps sufficient to note here that the best ISO case for an ULIRG dominated by an AGN, Mrk 273, does not differ much, for instance, from UGC 5101 or Arp 220 in global observational properties. Its double nucleus morphology is similar to Arp 220, its molecular content is very similar to both others observed recently with high resolution, mm interferometry (Downes and Solomon 1997, Scoville, Yun and Bryant 1997, this paper), its ratio of (ROSAT) X-ray luminosity to infrared luminosity is the same as those of the others within a factor of 2 (Rigopoulou et al. 1996) and its optical spectrum is that of a LINER or Seyfert 2 (Sanders et al. 1988a, Veron-Cetty and Veron 1993, UGC 5101 is a Seyfert 1.5 !). Recent ASCA observations show a moderately absorbed ($N(H) \sim 10^{23.4}$ cm$^{-2}$) hard X-ray source with $L_X/L_{IR} \sim 10^{-3}$ (Kii et al. 1997) which is a factor of ten higher than in Arp 220 but still far away from Seyfert 1 or 2 templates.

We now want to investigate the impact of our ISO results on evolutionary scenarios for ULIRGs. For this purpose we have compared the AGN 'activity level' (from 0 to 100% of the bolometric luminosity), as determined by the new ISO diagnostic tools, with the evolutionary state of the merger (assuming that all ULIRGs are mergers), as indicated by the separation of the two nuclei (or upper limit of it). For this analysis we included arcsecond or better resolution, optical/near-IR or radio imaging data from Sanders et al.(1988a,b), Graham et al.(1990), Carico et al.(1990), Condon et al.(1991 b), Sopp and Alexander (1991) and Majewski et al.(1993) for most ULIRGs in Table 2. To improve the AGN statistics we have added to the ISO sample several template AGNs from the IRAS catalogue that have ULIRG luminosities (IZw 1, Mrk 463, Pks1345+12). The resulting distributions for 6 starburst dominated and 6 AGN dominated ULIRGs are listed in Table 4. In the merger model of ULIRGs one would expect the two nuclei in the merger system to be the closer the later the phase of the merger. In the model of Sanders et al. (1988a) the more advanced mergers are dominated by a buried QSO. However, ***there is no such obvious correlation in the dozen ULIRGs with imaging data from the references listed above***. Arp 220 and UGC 5101 are the two most compact ULIRGs (nuclei separation d≤0.5 kpc) in the sample but appear to be dominated by star formation. The AGN dominated ULIRG Mrk 273 has a nuclear separation of ~ 0.7 kpc. Other ULIRGs dominated by star formation (e.g. NGC 6240, IRAS 22491, IRAS 12112, IRAS 14348) have nuclear separations between 0.6 and 6 kpc. Likewise for AGNs with $L≥10^{12}$ $L_o$ nuclear separations (if available) range from less than 0.5 kpc (IZw 1) to 4-8 kpc (Mrk 463, Pks1345+12). Clearly this important test ought to be re-done on a larger unbiased sample with high quality, 0.1-0.2" resolution, near-infrared imaging with the NICMOS camera onboard HST and/or with ground-based, adaptive optics. Based on the presently available data ***there is no obvious trend for the AGN component in ULIRGs to dominate in the more advanced mergers of our sample***. This then implies that the differences of the globally dominating luminosity sources must be controlled by local effects and that they are probably evolving on dynamical time scales. Relevant effects include the accretion rate onto the central massive black hole, the radiation efficiency of the accretion flow and the compression of the interstellar gas on ~1 kpc scale.

A luminosity of $~10^{12}$ $L_o$ requires a central massive black hole of mass $≥3x10^7$ $M_o$ and an accretion rate of $≥0.5/ε_{0.1}$ $M_o$ $yr^{-1}$ where $ε_{0.1}$ is the radiation efficiency of the accreting material in units of 10%. These numbers clearly show that ULIRG luminosities do require both fairly massive central black holes *as well as* efficient radiation and accretion. Investigations of massive black holes in nearby galaxies show that central black holes of mass a few $10^6$ to $10^8$ $M_o$ are probably common in galaxies but that low accretion rates or low radiation efficiency, or both, appear to be the norm rather than the exception (Genzel, Hollenbach and Townes 1994, Kormendy and Richstone 1995, Myoshi et al. 1995, Greenhill et al. 1995, Eckart and Genzel 1996, 1997). A plausible conclusion from our ISO observations may be that the accretion rate onto the existing central hole and the associated radiation efficiency of the accretion flow are the key factors in determining whether AGN or circum-nuclear star formation dominates the luminosity of ULIRGs. In addition, during the active ULIRG phase of $10^7$ to $10^8$ years the mass of the central hole is expected to grow substantially if the accretion rate is $\sim 10^0$ $M_o$ $yr^{-1}$. The presence of circum-nuclear gas and star formation is a sine qua non for the possibility of any substantial accretion onto the hole. However, the actual accretion rate onto the central black hole likely is controlled by the physics in its immediate vicinity (angular momentum transport of central gas streamers, wind pressure of nuclear massive stars etc.).

The burst ages and durations we estimate are several tens of million years. These time scales are much shorter than, for instance, the time scales of a few hundred million years that are required to reach the present stages of an advanced merger in most of the ULIRGs. Moreover, all of the ULIRGs studied with CO mm-spectroscopy seem to be (still) rich in molecular material, implying that there is enough 'fuel' to continue forming stars at the present rate for another $\sim 10^8$ years or so. It is therefore unclear why most of the galaxies look like ageing bursts with rapidly decaying star formation activity. A possible clue may come from recent spatially resolved observations of nearby starburst templates, such as M82 (Rieke et al. 1993, Satyapal et al. 1997, Förster-Schreiber et al. 1997) or IC 342 (Böker et al. 1997). It appears that in these systems burst activity starts at the nuclei but burns out fairly rapidly, presumably due to the destructive effects of stellar winds and supernovae. Star formation activity then moves outward from the nucleus but is happening there at present at a level that is more than an order of magnitude smaller than during the peak nuclear burst. Globally, then, M82 appears as a starburst with a lifetime of ~10 million years, or only a few times the lifetime of a single, moderately massive O star. It is, of course, not clear whether this scenario of 'a burst once begun rapidly terminates in a few O-star lifetimes' can be transferred to the larger and much more massive ultra-luminous galaxies. However, the powerful superwinds in ULIRGs found by Heckman et al. (1990) and Armus et al.(1990) do present strong evidence for the global destructive effects of star formation in these galaxies. If the mass of gas available and the rate of star formation do not determine the ages of the bursts what does ? Perhaps the N-body simulations of galaxy mergers by Barnes and Hernquist (1996) and Mihos and Hernquist (1996) are a guide for answering this important question. These models show that very strong compression of the interstellar gas occurs during/after each of a few (2 to 3) peri-center passages happening during the merger process and prior to the final merging of the two nuclei. This compression occurs as a result of strong gravitational torques removing a large fraction of the angular momentum of the gas and leading to fast inflow into the two nuclei. In the last stage of the merger the gas collects in the center of the merger remnant, between the two original nuclei. If star formation follows a Schmidt law ($R_* \sim (\rho_{gas})^n$, n=1 ... 2) the star formation history is dictated by several brief compression episodes at pericenter. Since most of the double nuclei in ULIRGs have separations of $\leq 1$ kpc, the dynamical time scale, and thus the separation between compressions, is several ten million years, consistent with the burst ages we have estimated above.

# 7. CONCLUSIONS

As part of the MPE ISO Central Program project 'galactic nuclei' we have carried out an ISO SWS and ISOPHOT-S spectroscopic survey of fifteen ultra-luminous IRAS galaxies (L~$10^{11.8}$ to $10^{12.5}$ $L_o$), as well as of about thirty starburst and AGN templates. As a detailed case study we also have observed one of the ultra-luminous galaxies, UGC 5101, with arcsecond resolution near-infrared and millimeter imaging spectroscopy.

Recent massive star formation plays an important or even dominant role in accounting for the bolometric luminosity of most (but not all) ULIRGs we have studied. The ratio of high to low excitation mid-IR nebular lines and the strength of the 6-11µm PAH feature indicate that only three of fifteen objects (Mrk 273, IRAS 23060+0505, Mrk 231) are likely dominated by an active nucleus. For the majority of our sources the new mid-IR excitation diagnostics as well as the Lyman continuum luminosity derived from low excitation IR nebular lines suggest that more than half of their $\geq 10^{12}$ $L_o$ bolometric luminosity comes from a starburst of age $10^7$ to $10^8$ years.

We confirm earlier findings that ultra-luminous IRAS galaxies often are composite objects. Our case study suggests that UGC 5101 in this sense may be a fairly typical case. In that galaxy a compact nuclear, hot dust source is surrounded by a rapidly rotating, ~1 kpc radius disk or ring of stars, dense neutral and ionized interstellar gas. About half of the disk's mass of $3 \times 10^{10}$ $M_o$ is in form of molecular gas. The circum-nuclear disk is the location of a powerful starburst that has been ongoing for a few tens of million years and which provides the majority of UGC 5101's 8-1000µm luminosity. The hot dust source, a compact radio source associated with it and the Seyfert 1.5 optical emission line spectrum all signal the presence of a luminous active nucleus.

The ISO spectroscopy shows that dust obscuration of the infrared emission line regions is large (equivalent 'mixed case' extinctions of $A_V \sim 50$ to 1000), in agreement with the large concentrations of molecular gas there. A quantitative assessment of the source(s) of luminosity in ultra-luminous galaxies thus can only be made at mid- to far-IR wavelengths, or at hard X-rays.

Finally, the simultaneous presence of active star formation and AGNs in several of the galaxies we have studied demonstrates that both processes can occur in the same phase of the evolution of luminous infrared galaxies. In the sample of sources that we have observed with ISO we do not find a trend of increasing importance of the AGN component with merger evolution. Instead at any given time during the merger process the compression of the circum-nuclear interstellar medium and the local accretion rate onto the central black hole are responsible for controlling whether the starburst or the AGN component dominates the overall energy output.

*Acknowledgements.* We thank J.Gallimore, A.Krabbe, H.Kroker and R.Maiolino for help with the 3D observations and E.Schinnerer for the calibration of the millimeter interferometry. We are grateful to Th.Boller for discussions on X-ray observations of IRAS galaxies, to D.Downes and P.Solomon for communicating to us prior to publication their IRAM mm-interferometry results, to T.Nakagawa for communicating prior to publication recent ASCA hard X-ray results on ULIRGs, and to A.Pauldrach and R.-P. Kudritzki for communicating to us prior to publication their new non-LTE stellar atmospheres. We thank the anonymous referee for his comments. SWS and the ISO Spectrometer Data Center at MPE are supported by DARA grants 50 QI 8610 8 and 50 QI 9402. This work has also been supported in part by the German-Israeli-Foundation grant I-196-137.7/91.

# Appendix A. Gas mass and CO to $H_2$ conversion factor in UGC 5101

In this section the dynamical measurements of the molecular gas presented in section 4.4 (Fig.12) to assess the questions of how much of the mass is due to stars versus gas, and what the conversion factor from CO intensity to $H_2$ column density is in UGC 5101. We take the dynamical mass within R~2" estimated above and subtract from it the mass of the old and young stellar components. Taking our own Galaxy as typical for average spiral galaxies the mass of the (old) stellar bulge within 1.5 kpc of the nucleus may range between 1 and $1.3 \times 10^{10}$ $M_o$. For a bolometric luminosity of $10^{12}$ $L_o$, a dereddened K-band stellar luminosity of $L_K = 5$ to $8 \times 10^9$ $L_o$ and the starburst parameters of UGC 5101 determined below (t~$\Delta$t~ 1 to $3 \times 10^7$ years, $M_*/L_K$~0.25 to 0.5) the mass of the young stellar component probably ranges between 1.5 and $5 \times 10^9$ $M_o$, depending on the shape of the initial mass function (Salpeter vs. Scalo) and the lower mass cutoff (the range above includes $m_l$ from 0.3 to 1 $M_o$). The combined old and young stellar mass components within the central 1.5 kpc of the nucleus thus add up to a mass of $1.5 \pm 0.3 \times 10^{10}$ $M_o$, leaving a mass of about $1.5 \times 10^{10}$ $M_o$ for the contribution from interstellar gas. Increasing the inclination from the most probable value (i=$60^o$) to the largest possible one ($90^o$) would decrease that mass to $0.8 \times 10^{10}$ $M_o$. Decreasing the inclination to $45^o$ would increase the mass to $3.2 \times 10^{10}$ $M_o$. *The circum-nuclear gas concentration in UGC 5101 thus contributes at least one third, at most two thirds and most likely about half of the total dynamic mass*, in agreement with similar conclusions in other ULIRGs (Solomon et al. 1997, Scoville et al. 1991, 1997). From our interferometer data in Fig.12 we find a total flux of the CO 1-0 emission within R=2" of ~50 Jy km/s. This corresponds to a total molecular mass (including 38% of mass in helium) of $1.6 \times 10^{10} (X/X_G)$ $M_o$ where $X_G = (N(H_2)/I(CO\ 1-0))_G = 2.3 \times 10^{20}$ $cm^{-2}$/(K km/s) is the Galactic conversion factor from CO 1-0 flux to $H_2$ column density (Bloemen et al. 1989). That value is in remarkable agreement with the gas mass we have estimated above from the total dynamical mass within the outer boundaries of the starburst ring, indicating that $X_{UGC5101}/X_G = 10^{0.0 \pm 0.3}$. Solomon et al.(1997) and Downes and Solomon (1997) have concluded that for the unusual physical conditions of the neutral ISM in ultra-luminous galaxies (no self-gravitating clouds, warm medium) the conversion factor should be below that in our Galaxy: $X_{ULIRG} \sim 0.3 \, X_G$. In the specific case of Arp 220 Scoville, Yun and Bryant(1997) find a best fit between CO kinematics and CO flux distribution for $X_{Arp\,220} \sim 0.6 \, X_G$ while Downes and Solomon (1997) find $X_{Arp\,220} \sim 0.3 \, X_G$. From a similar analysis as ours in the luminous Seyfert 1/starburst galaxy NGC 7469 Tacconi et al. (1997) deduce $X_{NGC\,7469} \sim X_G$. In contrast Shier et al. (1994) find $X \sim 0.1 \, X_G$ for several luminous IR galaxies (including Arp 220) from an analysis of the near-infrared stellar dynamics. However, Scoville, Yun and Bryant(1997) have correctly pointed out that in Arp 220 the stars - like the gas - should follow the dynamics of an inclined, rotating disk rather than an isotropic spherical distribution as assumed by Shier et al.(1994). The conversion factor of Shier et al.(1994) would then be a lower limit. Given the rather substantial uncertainties in all the parameters it is not clear whether the apparent differences between the derived X values are real or just reflect the uncertainties in the different interpretations. In the latter case it may be appropriate to average estimates from different authors. In summary then the different estimates of the conversion factor from $^{12}CO$ 1-0 intensity to $H_2$ column density in ULIRGs suggest an overall mean value somewhat below but near the Galactic one ($X_{ULIRG} \sim 0.5$ to $0.8 \, X_G$).

# Table 1

## ISO-SWS and ISO-PHOT-S Data on ULIRGs

| (1) SOURCE | (2) D (Mpc) | (3) Br β W/cm2 | (4) Br α W/cm2 | (5) [NeII] 12 W/cm2 | (6) [NeV] 14 W/cm2 | (7) [SIII] 18 W/cm2 | (8) [OIV] 24 W/cm2 | (9) [SIII] 33 W/cm2 | (10) [SiII] 34 W/cm2 | (11) PAH strength |
|---|---|---|---|---|---|---|---|---|---|---|
| UGC 5101 | 158 | ≤9.0E-21 | | | ≤1.5E-20 | ≤1.4E-20 | ≤6.0E-21 | 2.5E-20 | 6.6E-20 | 2.9 |
| 12112+0305 | 291 | ≤4.0e-21 | | | ≤6.0E-21 | ~8.e-21 | ≤8.0E-21 | ~1.5E-20 | ≤2.0e-20 | 2.4 |
| Mk 231 | 169 | ≤3.0e-21 | | ≤1.6E-20 | ≤1.5E-20 | ≤3.0E-20 | ≤3.0E-20 | ≤3.0E-20 | ~5.E-20 | 0.4 |
| Mk 273 | 151 | 4.8E-21 | | 3.0e-20 | 8.2E-21 | ≤8.2E-21 | 2.8E-20 | 2.3E-20 | 3.0E-20 | 1.9 |
| 14348-1447 | 330 | 1.5E-21 | | 1.7E-20 | ≤7.0E-21 | ≤5.0E-21 | ≤4.0E-21 | ~1.E-20 | 1.1E-20 | 2.5 |
| 15250+3609 | 213 | ≤1.0e-20 | | | ≤2.0E-21 | ≤7.0E-21 | ≤4.0E-21 | ≤1.2E-20 | ≤1.2E-20 | 3.1 |
| Arp 220[a] | 73 | 3.5E-21 | 1.8E-20 | 5.4E-20 | ≤3.5E-21 | ≤8.0e-21 | ≤4.0E-21 | 5.3E-20 | 5.6E-20 | 5.1 |
| N 6240 | 97 | ≤1.1e-20 | | 1.8E-19 | ≤1.0E-20 | ≤4.0E-20 | 3.1E-20 | 4.5E-20 | 2.1E-19 | 2.2 |
| 17208-0014 | 170 | ≤1.0e-20 | | 2.0e-20 | ≤6.5e-21 | ≤1.0e-20 | ≤7.0e-21 | 3.6e-20 | 4.0e-20 | 5.8 |
| 19254-7245 | 246 | ≤1.2e-20 | | ≤8.0E-20 | ≤4.0E-20 | ≤3.0E-20 | ≤2.5E-20 | ≤1.5E-19 | ≤2.0E-19 | 0.8 |
| 20100-4156 | 517 | | | 6.2E-21 | ≤6.0E-21 | ≤2.3E-21 | ≤7.8E-21 | 9.3E-21 | ~8.E-21 | 2.7 |
| 20551-4250 | 171 | ≤3.8E-21 | | 1.3E-20 | ≤2.5E-21 | 3.0E-21 | ≤3.0E-21 | 1.4E-20 | 1.7E-20 | 3.0 |
| 22491-1808 | 309 | ≤4.0e-21 | | 7.1e-21 | ≤5.0e-21 | ≤5.0e-21 | ≤1.0e-20 | ≤1.5e-20 | ≤2.5e-20 | 1.5 |
| 23060+0505 | 692 | | | | | | | | | ≤0.4 |
| 23128-5919 | 178 | ≤3.5E-21 | | 5.6E-20 | ≤4.0E-21 | 8.9E-21 | 3.0E-21 | 2.8E-20 | 3.7E-20 | 1.5 |

The organization of Table 1 is as follows.

Column 1: source name

Column 2: distance derived for $H_o$=75 km/s/Mpc and $q_o$=0.5

Column 3 to 10: observed ISO integrated line fluxes in W cm$^{-2}$. Wherever limits are given, these are derived from the 'most probable' error estimates (~3-5σ) discussed in section 2. Absolute flux uncertainties are ±25%.

Column 11: strength of the 7.7μm PAH feature. This we define to be the ratio of the peak 7.7μm flux to the 7.7μm continuum as obtained from a linear interpolation between 5.9 and 11.0μm which are both outside of the wavelength regions affected by PAH emission yet close enough to the 7.7μm feature that a simple linear interpolation of the continuum is a sensible first order approximation (even when a significant 9.7μm silicate absorption is present).

(a) in addition Pfα : $4.4 \times 10^{-21}$ W cm$^{-2}$. In comparison to Sturm et al.(1996), the line fluxes presented here (containing new data and a better calibration) agree within 10%, with the exception of 33μm [SIII] and 34μm [SiII] which are now almost a factor of two lower due to addition of new data and replacement of the preliminary calibration file in the longest wavelength band of SWS that was used by Sturm et al. (1996).

# Table 2. Derived Properties of Galaxies observed with ISO[a]

| (1) | (2) | (3) | (4) | (5) | (6) | (7) | (8) | (9) | (10) | (11) | (12) | (13) | (14) | (15) | (16) | (17) |
|---|---|---|---|---|---|---|---|---|---|---|---|---|---|---|---|---|
| SOURCE | D (Mpc) | IRAS 12.0 | IRAS 25.0 | IRAS 60.0 | IRAS 100.0 | L(FIR) (Lo) | A(V) (scrn) | A(V) (mxd) | FIR/ Lyc(rec) | FIR/ 64[NeII] | FIR/ 90[SIII] | FIR/ 75[SiII] | <FIR/ L,Ne,S,Si> | PAH strength | [OIV]/[NeII] or 1.7[SIII] | [NeV]/[NeII] or 1.7[SIII] |
| **STARBURST TEMPLATES** | | | | | | | | | | | | | | | | |
| N253 | 3 | 24.0 | 120.0 | 784.0 | 993.0 | 1.2E+10 |  | 30 | 21 | 29 |  | 29 | 26 | 3.1 | ≤0.017 | ≤0.032 |
| N3256 | 37 | 3.2 | 16.0 | 88.3 | 115.3 | 2.5E+11 |  | 35 | 7 | 6 | 6 | 5 | 6 | 4.2 | 0.002 | ≤0.012 |
| M82 | 3 | 53.2 | 274.0 | 1168.0 | 1145.0 | 2.5E+10 | 10 |  | 12 | 10 | 10 | 9 | 10 | 5.4 | 0.008 | ≤0.001 |
| N3690A | 40 | 3.7 | 21.6 | 105.0 | 110.0 | 3.7E+11 |  | 20 | 24 | 32 | 28 | 23 | 27 | 5.0 | ≤0.011 | ≤0.02 |
| N3690B | 40 | 3.7 | 21.6 | 105.0 | 110.0 | 2.3E+11 |  | 20 | 13 | 22 | 13 | 16 | 16 | 4.0 | ≤0.013 | ≤0.022 |
| N4038/39 | 21 | 1.2 | 3.9 | 39.2 | 76.0 | 3.7E+10 |  | 80 | 8 | 17 | 7 | 12 | 11 | 2.3 | ≤0.013 |  |
| N4945 | 4 | 4.0 | 14.5 | 359.0 | 621.0 | 1.0E+10 |  | 100 | 19 | 11 | 32 |  | 21 | 10.7 | ≤0.008 | ≤0.008 |
| M83 | 5 | 4.8 | 19.6 | 110.3 | 236.0 | 1.1E+10 | 5 |  | 23 | 9 | 13 | 11 | 14 | 2.8 | ≤0.005 | ≤0.006 |
| N5253 | 4 | 2.6 | 12.0 | 31.0 | 29.4 | 8.0E+08 | 15 |  | 2 | 24 | 5 | 11 | 11 | 0.5 | 0.021 | ≤0.015 |
| G.C. | 8.E-03 |  |  |  |  | 1.5E+07 | 32 |  | 8 | 8 | 22 | 28 | 17 | 3.8[a] | 0.003 | ≤0.008 |
| N7469 | 66 | 1.3 | 5.8 | 25.9 | 34.9 | 3.0E+11 |  | 20 | 30 | 7 |  |  | 18 | 1.1 | 0.051 | ≤0.018 |
| N7552 | 21 | 3.0 | 12.2 | 72.0 | 102.0 | 9.8E+10 | 5 |  | 29 | 18 | 21 | 15 | 21 | 2.7 | ≤0.01 |  |
| **ULIRGs** | | | | | | | | | | | | | | | | |
| UGC5101 | 158 | 0.3 | 1.0 | 11.5 | 20.2 | 8.3E+11 |  | 50[b] | (33)[c] |  | 36 | 16 | 28 | 2.9 | ≤0.17 | ≤0.6 |
| 12112+0305 | 291 | 0.1 | 0.5 | 8.5 | 10.0 | 1.7E+12 | ~5 |  |  |  | 45 | 40 | 42 | 2.4 | ≤0.3 | ≤0.27 |
| Mrk231 | 169 | 1.9 | 8.7 | 32.0 | 30.3 | 3.0E+12 | [≥5] |  |  |  |  | 81 | 81 | 0.4 | ≤0.6 | ≤0.33 |
| Mrk273 | 151 | 0.2 | 2.3 | 21.7 | 21.4 | 1.2E+12 | ≥15 |  | 15 | 50 | 68 | 63 | 49 | 1.9 | 0.85 | 0.27 |
| 14348-1447 | 330 | 0.1 | 0.5 | 6.9 | 7.1 | 1.8E+12 | ≥5 |  | 30 | 40 | 55 | 60 | 46 | 2.5 | ≤0.21 | ≤0.41 |
| 15250+3609 | 213 | 0.2 | 1.3 | 7.3 | 5.9 | 8.8E+11 |  |  |  |  |  |  |  | 3.1 |  |  |
| Arp220 | 73 | 0.5 | 7.9 | 104.0 | 112.0 | 1.3E+12 | 45 | ~15 |  | 43 | 99 | 112 | 37[d] | 5.1 | ≤0.031 | ≤0.06 |
| N6240 | 97 | 0.6 | 3.4 | 22.7 | 27.8 | 6.0E+11 | ≥5 |  | 15 | 47 | 12 | 25 |  | 2.2 | 0.16 | ≤0.05 |
| 17208-0014 | 170 | 0.2 | 1.7 | 31.1 | 34.9 | 2.1E+12 | ≥25 |  |  | 72 | 54 | 58 | 61 | 5.8 | ≤0.14 | ≤0.19 |
| 19254-7245 | 246 | 0.2 | 1.2 | 5.5 | 5.8 | 1.0E+12 |  |  |  |  |  |  |  | 0.8 |  |  |
| 20100-4156 | 517 | 0.1 | 0.3 | 5.2 | 5.2 | 3.3E+12 | ≥30 |  |  | 33 | 34 | 47 | 38 | 2.7 | ≤0.62 | ≤0.78 |
| 20551-4250 | 171 | 0.3 | 1.9 | 12.8 | 10.0 | 9.5E+11 | 35 |  |  | 34 | 56 | 55 | 48 | 3.0 | ≤0.12 | ≤0.18 |
| 22491-1808 | 309 | 0.1 | 0.6 | 5.4 | 4.5 | 1.2E+12 | [20] |  |  | 42 | 24 | 17 | 28 | 1.5 | ≤0.45 | ≤0.32 |
| 23060+0505 | 692 | 0.2 | 0.4 | 1.2 | 0.8 | 2.4E+12 |  |  |  |  |  |  |  | ≤0.37 |  |  |
| 23128-5919 | 178 | 0.2 | 1.6 | 10.8 | 11.0 | 9.1E+11 | 20 |  |  | 12 | 29 | 26 | 23 | 1.5 | 0.037 | ≤0.07 |
| **AGN TEMPLATES** | | | | | | | | | | | | | | | | |
| IZw 1 | 242 | 0.5 | 1.2 | 2.2 | 2.6 | 7.1E+11 |  |  |  |  |  |  |  | ≤0.18 |  |  |
| Mrk1 | 65 | 1.9 | 0.9 | 2.5 | 2.9 | 9.5E+10 |  |  |  |  |  |  |  | 1.8 |  |  |
| N1068 | 14 | 38.0 | 86.0 | 186.0 | 239.0 | 3.0E+11 | 8 |  | 56 | 111 | 99 | 62 | 82 | 0.27 | 2.3 | 1.1 |
| M81 | 4 | 0.8 | 0.8 | 6.8 | 32.0 | 1.0E+08 |  |  |  |  |  |  |  | 0.18 |  |  |
| N3783 | 34 | 0.8 | 2.5 | 3.3 | 4.9 | 2.5E+10 | 0 |  |  |  | 15 | 9 | 12 | ≤0.19 | 0.4 | 0.24 |
| N4151 | 20 | 1.8 | 4.7 | 6.9 | 10.2 | 1.7E+10 | 3 |  | 24 | 10 | 15 | 8 | 14 | 0.06 | 1.6 | 0.5 |
| 3C273 | 632 | 0.6 | 0.9 | 2.1 | 2.9 | 4.6E+12 |  |  |  |  |  |  |  | ≤0.2 |  |  |
| M87 | 17 | 0.2 | 0.2 | 0.4 | 1.0 | 1.0E+09 |  |  |  |  |  |  |  | ≤0.4 |  |  |
| N4569 | 15 | 0.5 | 1.1 | 7.6 | 23.7 | 7.0E+09 |  |  |  |  |  |  |  | 1.45 |  |  |
| Circinus | 3 | 18.9 | 68.4 | 248.0 | 314.0 | 7.0E+09 | 20 |  | 29 | 18 | 18 | 13 | 20 | 1.7 | 0.4 | 0.4 |
| Cen A | 4 | 13.3 | 17.3 | 162.0 | 314.0 | 4.0E+09 | 30 |  | 19 | 15 | 23 | 11 | 17 | 1.3 | 0.24 | 0.13 |
| Mrk463 | 199 | 0.5 | 1.6 | 2.2 | 1.9 | 5.1E+11 |  |  |  |  |  |  |  | 0.16 |  |  |
| N5506 | 23 | 1.3 | 3.6 | 8.4 | 8.9 | 2.0E+10 | 8 |  | 83 | 28 | 17 | 7 | 34 | 0.4 | 1.3 | 0.3 |
| N5643 | 16 | 1.1 | 3.7 | 19.5 | 38.2 | 1.8E+10 | 1 |  | 75 |  | 46 | 52 | 57 | 0.67 | 1.0 |  |
| Pks2048-57 | 45 | 1.1 | 3.9 | 5.3 | 4.2 | 6.1E+10 | 0 |  |  |  | 21 | 18 | 20 | 0.19 | 1.0 |  |
| N7582 | 20 | 1.4 | 6.3 | 48.0 | 72.8 | 5.0E+10 | 18 |  |  | 20 | 24 | 12 | 19 | 2.5 | 0.67 | 0.22 |

Table 2 is organized as follows.

Column 1: source name

Column 2: distance [for $H_o$=75 km/s/Mpc and $q_o$=0.5]

Columns 3-6: IRAS flux densities (Jy)

Column 7: 8-1000 μm luminosity (=$L_{IR}$~$L_{bol}$ ) for the ISO (20-30") beam, using the prescription given in Table 1 of Sanders and Mirabel (1996): L(8-1000μm)=5.6x10$^5$ $D_{Mpc}^2$ ( $S_{100}$ + 2.58 $S_{60}$ + 5.16 $S_{25}$ + 13.48 $S_{12}$) where S is the flux density in Jy.

Columns 8 and 9: screen and mixed case extinctions in V-band, deduced from the ISO SWS data as described in section 3.1, or whenever that was not possible, using near-infrared data or adopted values (square brackets).

Columns 10 to 13: Various estimates of the ratio $L_{IR}$/$L_{Lyc}$ (uncertainties ±50% to a factor of 2). In column 10 the Lyman continuum luminosity $L_{Lyc}$ (hν≥13.6 eV) is derived from the extinction corrected Brα, Brβ or Brγ recombination lines ($T_e$=7500 K, $n_e$=300 cm$^{-3}$, case B, <hν$_{Lyc}$>~16 eV, ionization bounded, see section 3.6). Note that several of these assumptions are inappropriate for AGNs so that the values of $L_{IR}$/$L_{Lyc}$ obtained in this way are probably misleading for such sources (see footnote 5). In columns 11 to 13 $L_{Lyc}$ is derived from extinction corrected fine structure line luminosities tracing the ionized gas. These line luminosities can be assumed to be proportional to Lyman continuum luminosity with roughly a constant factor. This factor is obtained empirically from the average $L_{Lyc}$/L(fine structure line) ratio in starburst galaxies (see section 3.6). In column 11 we use the 12.8μm [NeII] line multiplied 64. In column 12 we use the 33.5μm [SIII] line luminosity multiplies by 90 and in column 13 we use the 34.8μm [SiII] line luminosity multiplied 75.

Column 14: Average $L_{IR}$/$L_{Lyc}$ ratio as determined from a mean (or median) of columns 10 to 13. As noted above the $L_{IR}$/$L_{Lyc}$ values listed for AGNs are probably misleading.

Column 15: Strength of the 7.7μm PAH feature (see explanation of column 11 in Table 1).

Column 16: Extinction corrected 25.9μm [OIV]/ 12.8μm [NeII] line ratio (or 3σ upper limit) or, if [NeII] is not available, [NeII] replaced by 1.7 times 33.5μm [SIII]. The factor 1.7 is the average [NeII]/[SIII] ratio in starburst galaxies. Line ratios are probably uncertain to ±30%.

Column 17: Extinction corrected 14.3μm [NeV]/ 12.8μm [NeII] line ratio (or 3σ upper limit) or, if [NeII] is not available, [NeII] replaced by 1.7 times 33.5μm [SIII]. Line ratios are probably uncertain to ±30%.

(a) molecular ring.

(b) derived from ISO [SIII] ratio, plus Hβ/Hα/Pα/Brγ/Brβ line ratios and J-H-K stellar colors. The equivalent screen

extinction is about 5 to 7.

(c) from Pα/Brγ.

(d) median of columns 10 through 12, plus the $L_{FIR}/L_{Lyc}$ values from the free-free 100GHz continuum (19) and from

the H92α recombination line (see footnote 8).

-------------------------------------------------------------------------------

| | |
|---|---|
| size of starburst zone | $R_{inner} \sim 200$ pc |
| | $R_{peak} \sim 420$ pc |
| | $R_{outer} \sim 1.5$ kpc |
| $L_{bol} \sim L_{IR}$ | $10^{11.92}$ $L_o$ |
| $L_K$(stellar, dereddened) | $10^{9.7}$ $L_o$ |
| $L_{Lyc}$ | $10^{10.6}$ $L_o$ |
| $M_{dyn}$ | $10^{9.96}$ $(R/380$ pc$)^{0.88}$ $M_o$ |
| $M_*$ (old, $R \leq 1.5$ kpc) | $10^{10.1}$ $M_o$ |
| $M_*$(young, $R \leq 1.5$ kpc) | $10^{9.4}$ $M_o$ |
| $M_{gas}(R \leq 1.5$ kpc) | $10^{10.18}$ $M_o$ |
| $< N(H_2) >$ | $10^{23.3 \pm 0.3}$ cm$^{-2}$ |
| $A_V$ (CO) | 190 |
| $A_V$(IR) | 5 to 7 (screen), 50 (mixed) |
| $t_{Burst} \sim \Delta t_{Burst}$ | 1.2 to $3 \times 10^7$ years |
| present star formation rate | $10^{1.7}$ $M_o$/yr ($m_l = 1$ $M_o$, Salpeter) |
| supernova rate | 1.8 yr$^{-1}$ |

**Table 4**

**Distribution of Nuclear Separations in ULIRGs**

| Nuclear Separation (kpc) | Number of Starburst Dominated ULIRGs | Number of AGN Dominated ULIRGs |
|---|---|---|
| d≤0.5 | 2 | 1 to 2 |
| 0.5≤d≤1.5 | 1 | 1 |
| 1.5≤d≤3 | 1 | 0 to 1 |
| 3≤d≤6 | 2 | 2 |
| 6≤d | | 1 |

# REFERENCES

Alexander, T. et al. 1998, in prep.

Allamandola, L.J., Sandford, S.A., Hudgins, D.M. and Witteborn, F.C. 1995, in From Gas To Dust, Eds. M.R.Haas, J.A.Davidson and E.F.Erickson, ASP Conference series (San Francisco) 73, 23

Antonucci, R. 1993, Ann.Rev.Astr.Ap. 31, 473

Armus, L., Heckman, T.M. and Miley, G.K. 1990, Ap.J. 364, 471

Armus, L., Neugebauer, G., Soifer, B.T. and Matthews, K. 1995, A.J.110, 610

Awaki, H., Kunieda, H., Tanaka, Y. and Koyama, K. 1991, Publ.Astr.Soc.Jap. 43, L37

Barnes, J.E. and Hernquist, L. 1996, Ap.J. 471, 115

Bloemen, H. 1989, Ann.Rev.Astr.Ap. 27, 469

Böker, T., Förster-Schreiber, N. and Genzel, R. 1997, Ap.J. submitted

Boller, Th., Bertoldi, F., Dennefeld, M. and Voges, W. 1997, Astr.Ap. in press

Brandt, W.N., Fabian, A.C., Takahashi, K., Fujimoto, R., Yamashita, A., Inoue, H. and Ogasaka, Y. 1997, MNRAS in press

Carico, D.P., Graham, J.R., Matthews, K., Wilson, T.D., Soifer, B.T., Neugebauer, G. and Sanders, D.B. 1990, Ap.J. 349, L39

Clements, D.L., Sutherland, W.J., McMahon, R.G. and Saunders, W. 1996b, MNRAS 279, 477

Clements, D.L., Sutherland, W.J., Saunders, W., Efstathiou, G.P., McMahon, R.G. et al. 1996a, MNRAS 279, 459

Condon, J.J., Anderson, M.L. and Helou, G. 1991a, Ap.J. 376, 95

Condon, J.J., Huang, Z.P., Yin, Q.F. and Thuan, T.X. 1991b, Ap.J. 378, 65

Contini, M. and Viegas, S.M. 1992, Ap.J. 401, 481

de Graauw, T. et al. 1996, Astr.Ap. 315, L49

Downes, D., Solomon, P.M. and Radford, S.J.E. 1993, Ap.J. 414, L13

Downes, D. and Solomon, P.M. 1997, in prep.

Doyon, R., Joseph, R.D. and Wright, G.S. 1994a, Ap.J. 421, 101

Doyon, R., Wells, M., Wright, G.S., Joseph, R.D., Nadeau, D. and James, P.A. 1994b, Ap.J. 437, L23

Draine, B. T. 1989, in Infrared Spectroscopy in Astronomy, ed.B.H.Kaldeich, ESA-SP, p93

Dudley, G.C. and Wynn-Williams, C.G. 1997, in press

Duley, W.W. and Williams, D.A. 1991, MNRAS 196, 269

Eckart, A. and Genzel, R. 1996, NATURE 383, 415

Eckart, A. and Genzel, R. 1997, MNRAS 284, 576

Egami, E. et al. 1997, in prep.

Ferland, G.J. 1993, Internal Report, Dep.of Physics, Univ. of Kentucky

Fischer, J. et al. 1996, Astr.Ap. 315, L 97

Förster-Schreiber, N. et al. 1997, in prep.

Furniss, I., Jennings, R.E. and Moorwood, A.F.M. 1974, in HII Regions and the Galactic Centre, 8[th] ESLAB symposium,  ESRO SP-105, 61

Gallimore, J.F., Baum, S.A. and O'Dea, C.P. 1997, NATURE in press

Genzel, R., Becklin, E.E., Moran, J.M., Reid, M.J., Jaffe, D.T., Downes, D. and Wynn-Williams, C.G. 1982, Ap.J. 255, 527

Genzel, R., Hollenbach, D.J. and Townes, C.H. 1994, Rep.Progr. Phys. 57, 417

Genzel, R., Weitzel, L., Tacconi-Garman, L.E., Blietz, M., Cameron, M. et al. 1995, Ap.J. 444, 129

Goldader, J.D., Joseph, R.D., Doyon, R. and Sanders, D.B. 1995, Ap.J. 444, 97

Goldader, J.D., Joseph, R.D., Doyon, R. and Sanders, D.B. 1997b, Ap.J.(Suppl) 108, 449

Goldader, J.D., Joseph, R.D., Sanders, D.B. and Doyon, R. 1997a, Ap.J. 474, 104

Graham, J.R., Carico, D.P., Matthews, K., Neugebauer, G., Soifer, B.T. and Wilson, T.D. 1990, Ap.J. 354, L5

Granato, G.L. and Danese, L. 1994, MNRAS 268, 235

Greenhill, L. et al. 1995, Ap.J. 440, 619

Guilloteau, S. et al. 1992, Astr.Ap. 262, 624

Heckman, T.M., Armus, L. and Miley, G. K. 1990, Ap.J. Suppl. 74, 833

Helou, G., Soifer, B.T. and Rowan-Robinson, M. 1985, Ap.J. 298, L7

Huang, Zh.P., Thuan, T.X., Chevalier, R.A., Condon, J.J. and Yin, Q.F. 1994, Ap.J. 424, 114

Kii, T., Nakagawa, T., Fujimoto, R., Ogasaka, T., Miyazaki, T., Kawabe, R. and Terashima, Y. 1997, in X-ray Imaging and Spectroscopy of Cosmic Hot Plasmas, eds. F.Makino and K.Mitsuda (Universal Academy Press:Tokyo), 161

Kim, D.C. 1995, Ph.D. Thesis, Univ.of Hawaii

Kessler, M.F. et al. 1996, Astr.Ap. 315, L 27

Kim, D.C., Sanders, D.B., Veilleux, S., Mazarella, J.M., and Soifer, B.T. 1995, Ap.J. Suppl. 98, 129

Kormendy, J. and Richstone, D. 1995, Ann.Rev.Astr.Ap. 33, 581

Kovo, O. and Sternberg, A. 1997, in prep

Krabbe, A., Sternberg, A. and Genzel, R. 1994, Ap.J. 425, 42.

Krabbe, A., Genzel, R., Eckart, A. et al. 1995, Ap.J. 447, L95

Kunze, D. et al. 1996, Astr.Ap. 315, L 101

Kunze, D. et al. 1997, in prep.

Kurucz, R.L. 1992, Rev.Mex.Astron.Astrofis. 23, 181

Lancon, A. and Rocca-Volmerange, B. 1996, New Astronomy 1, 215

Larkin, J.E., Armus, L., Knop, R.A., Mattewhs, K. and Soifer, B.T. 1995, Ap.J. 452, 599

Lemke, D. et al. 1996, Astr.Ap. 315, L 64

Lester, D.F. and Gaffney, N.I. 1994, Ap.J. 431, L13

Lonsdale, C.J., Smith, H.E. and Lonsdale, C.J. 1993, Ap.J. 405, L9

Lonsdale, C.J., Smith, H.E. and Lonsdale, C.J. 1995, Ap.J. 438, 632

Lonsdale, C.J., Lonsdale, C.J. and Smith, H.E. 1995, BAAS 187, 5008

Lord, S. 1992, NASA Technical Memo 103957

Lutz, D. et al. 1996a , Astr.Ap. 315, L 137

Lutz, D. et al. 1996b, Astr.Ap. 315, L 269

Lutz, D. et al. 1997, in prep.

Maiolino, R., Krabbe, A., Thatte, N. and Genzel, R. 1997, Ap.J. in press

Majewski, S.R., Hereld, M., Koo, D.C., Illingworth, G.D. and Heckman, T.M. 1993, Ap.J. 402, 125

Mattila, K. et al. 1996, Astr.Ap. 315, L353

Meynet, G. et al. 1994, Astr.Ap.(Suppl.) 103, 97

Mihos, J.C. and Hernquist, L. 1996, Ap.J. 464, 641

Mirabel, I.F., Lutz, D. and Maza, J. 1991, Astr.Ap. 243, 367

Mitsuda, K. 1995, in  Proc. Of 17[th] Texas Symposium, eds.  H.Böhringer, G.E.Morfill and
    J.Trümper, N.Y.Acad.Sci.(New York),213

Moorwood, A.F.M. 1996, Space Science Rev. 77, 303

Moorwood, A.F.M. and Oliva, E. 1988, Astr.Ap. 203, 278

Moorwood, A.F.M. and Oliva, E. 1994, Ap.J. 429, 602

Moorwood, A.F.M. et al. 1996, Astr.Ap. 315, L 109

Moorwood, A.F.M. et al. 1997, in prep.

Myoshi, M., Moran, J.M., Herrnstein, J., Greenhill, L., Nakai, N. et al. 1995, NATURE 373, 127

Netzer, H. 1990, in Active Galactic Nuclei, eds. T.J.-L.Courvoisier and M.Mayor, Springer (New
    York),57

Norman, C. 1991, in Massive Stars in Starbursts, ed. C.Leitherer, N.R. Wolborn, T.M. Heckman and
    C.A.Norman, (Cambridge: Cambridge Univ.Press), 271

Norman, C.A. and Scoville, N.Z. 1988, Ap.J. 332, 124

Oliva, E., Origlia, L., Kotilainen, J.K. and Moorwood, A.F.M. 1995, Astr.Ap. 301, 55

Origlia, L., Moorwood, A.F.M. and Oliva, E. 1993, Astr.Ap. 280, 536

Pauldrach, A. et al. 1997, in prep.

Pier, E.A. and Krolik, J.H. 1993, Ap.J. 418, 673

Puget, J.-L. and Leger, A. 1989, Ann.Rev.Astr.Ap. 27, 161

Rieke, G.H. 1988, Ap.J. 331, L5

Rieke, G.H., Cutri, R., Black, J.H., Kailey, W.F., McAlary, C.W., Lebofsky, M.J. and Elston, R. 1985, Ap.J. 290, 116

Rieke, G.H. , Loken, K., Rieke, M.J. and Tamblyn, P. 1993, Ap.J. 412, 99

Rigopoulou, D. et al. 1996, Astr.Ap. 315, L 125

Rigopoulou, D., Lawrence, A. and Rowan-Robinson, M. 1996, MNRAS 278, 1049

Rigopoulou, D. et al. 1997, in prep.

Roche, P.F., Aitken, D.K., Smith, C.H. and Ward, M.J. 1991, MNRAS 248, 606

Rowan-Robinson, M. and Crawford, J. 1989, MNRAS 238, 523

Rowan-Robinson, M. and Efstathiou, A. 1993, MNRAS 263, 675

Sanders, D.B. and Mirabel, I.F. 1996, Ann.Rev.Astr.Ap. 34, 749

Sanders, D.B., Egami, E., Lipari, S., Mirabel, I.F. and Soifer, B.T. 1995, A.J. 110, 1993

Sanders, D.B., Soifer, B.T., Elias, J.H., Madore, B.F., Matthews, K., Neugebauer, G. and Scoville, N.Z. 1988a, Ap.J. 325, 74

Sanders, D.B., Soifer, B.T., Elias, J.H., Neugebauer, G. and Matthews, K. 1988b, Ap.J. 328, L35

Satyapal, S., Watson, D.M., Pipher, J.L., Forrest, W.J., Greenhouse, M.A., Fischer, J., Woodward, C.E. and Smith, H.A. 1997, Ap.J. in press.

Schaerer, D. , Meynet, G., Maeder, A. and Schaller, G. 1993, Astr.Ap.(Suppl.) 102, 339

Scoville, N.Z. and Norman, C.A. 1988, Ap.J. 332, 163

Scoville, N.Z., Sargent, A.I., Sanders, D.B. and Soifer, B.T. 1991, Ap.J. 366, L5

Scoville, N.Z., Yun, M.S. and Bryant, P.M. 1997, Ap.J. submitted

Scoville, N.Z., Thompson, R. et al. 1998, in prep.

Shaya, E.J., Dowling, D.M., Currie, D.G., Faber, S.M. and Groth, E.J. 1994, A.J. 107, 1675

Shier, L.M., Rieke, M.J. and Rieke, G.H. 1994, Ap.J. 433, L9

Simpson, J.P., Colgan, S.W.J., Rubin, R.H., Erickson, E.F. and Haas, M.R. 1995, Ap.J. 444, 721

Smith, C.H., Aitken, D.K. and Roche, P.F. 1989, MNRAS 241, 425

Smith, H.E., Lonsdale, C.J., Lonsdale, C.J. and Diamond, P.J. 1997, preprint

Soifer, B.T., Houck, J.R. and Neugebauer, G. 1987, Ann.Rev.Astr.Ap. 25, 187

Soifer, B.T., Sanders, D.B., Neugebauer, G., Danielson, G.E., Lonsdale, C.J., Madore, B.F. and
    Persson, S.E. 1986, Ap.J. 303, L41

Soifer, B.Y. et al. 1984, Ap.J. 278, L71

Solomon. P.M., Donwes, D., Radford, S.J.E. and Barrett, J.W., 1997, Ap.J. 478, 144

Sopp, H.M. and Alexander, P. 1991, MNRAS 251, 112

Spinoglio, L. and Malkan, M.A. 1992, Ap.J. 399 , 504

Sternberg, A. and Dalgarno, A. 1989, Ap.J. 338, 199

Sturm, E. et al. 1996, Astr.Ap. 315, L 133

Sturm, E. et al. 1997, in prep.

Sutherland, R.S., Bicknell, G.V. and Dopita, M.A. 1993, Ap.J. 414, 510

Tacconi, L.J., Genzel, R., Blietz, M., Cameron, M., Harris, A.I. and Madden S. 1994, Ap.J. 426,
    L77

Tacconi, L.J., Gallimore, J., Genzel, R. and Tacconi-Garman, L.E. 1997, Ap.J. submitted

Tezca, M. et al. 1997, in prep.

Thatte, N., Kroker, H., Weitzel, L., Tacconi-Garman, L.E., Tecza, M., Krabbe, A. and Genzel, R.
    1995, Proc. SPIE, 2475, 228

Thatte, N. , Quirrenbach, A., Genzel, R. and Tecza, M. 1997, Ap.J. in press

Tielens, A. and Hollenbach, D.J. 1985, Ap.J. 291, 722

van der Hulst, J.M., Terlouw, J.P., Begeman, K., Zwitser, W. and Roelfsema, P.R. 1992, in
    Astronomical Data Analysis Software and Systems I, eds. D.M.Worall, C.Biemesderfer and
    J.Barnes, ASP Conf series 25, 131

van der Werf, P.P., Genzel, R., Krabbe, A., Blietz, M., Lutz, D., Drapatz, S., Ward, M.J. and
    Forbes, D.A. 1993, Ap.J. 405, 522

van Dishoeck, E.F. and Black, J.H. 1988, Ap.J. 334, 711

Veilleux, S., Kim, D.C., Sanders, D.B., Mazarella, J.M. and Soifer, B.T. 1995, Ap.J.(Suppl.) 98, 171

Veilleux, S., Sanders, D.B. and Kim, D.C. 1997, Ap.J. 484, 92

Veron-Cetty, M.-P. and Veron, P. 1993, ESO Scientific Report  No13 (ESO:Garching)

Verstraete, L. et al. 1996, Astr.Ap. 315, L337

Vigroux, L. et al. 1996, Astr.Ap. 315, L 93

Voit, G.M. 1992, Ap.J. 399, 495

Weitzel, L., Krabbe, A., Kroker, H., Thatte, N., Tacconi-Garman, L.E., Cameron, M. and Genzel, R.
    1996, Astr.Ap.(Suppl.) 119, 531

Wood, D. and Churchwell, E. 1989, Ap.J.Suppl. 69, 831

Young, S., Hough, J.H., Axon, D.J., Ward, M.J. and Bailey, J.A. 1996, MNRAS 280, 291

Zhao, J.H., Anantharamaiah, K.R., Goss, W.M. and Viallefond, F. 1996, Ap.J. 472, 54

# FIGURE CAPTIONS

Fig.1. ISO SWS spectra ($\lambda/\Delta\lambda\sim1500$) of ultra-luminous galaxies, ordered in terms of increasing red-shift. The 12.8$\mu$m [NeII], 14.3$\mu$m [NeV], 18.7$\mu$m [SIII], 25.9$\mu$m [OIV], 33.5$\mu$m [SIII] and 34.8$\mu$m [SiII] spectra are shown (where observed) for the fourteen ULIRGs in our SWS sample. Identifications and red-shifts are given to the left of the spectra and for each spectrum a velocity range of $\pm1000$ km/s (IRAS 19254: $\pm2000$ km/s) around the rest wavelength is shown. Flux densities are in Jy (absolute uncertainty $\sim25\%$). Typical 'most probable' $\pm1\sigma$ error bars, often dominated by various systematic effects (Sect. 2.1), are marked.

Fig.2. ISOPHOT-S spectra ($\lambda/\Delta\lambda\sim90$) of 15 ULIRGs. Observed wavelengths are shown at the top and rest wavelengths at the bottom of each spectrum. Flux densities (not including the correction for chopped mode of a factor of $\sim1.4$) are in Jy. Error bars as provided by PIA (Sect. 2.1) are shown for UGC5101.

Fig.3. ISO SWS line ratios in ultra-luminous galaxies, starburst templates and AGNs. A typical error bar is shown (for NGC 1068). Left: Extinction corrected 26$\mu$m [OIV]/12.8$\mu$m [NeII] (or alternatively 1.7x(33.4$\mu$m [SIII])) line ratios for starburst templates  (open triangles), ULIRGs (filled circles) and AGNs (crossed rectangles). Right: Same for 14$\mu$m [NeV]/12.8$\mu$m [NeII] (or alternatively 1.7x(33.4$\mu$m [SIII])).

Fig.4. 7.7µm 'PAH' feature strength in different sources. Plotted is the peak 7.7µm feature flux (in ISOPHOT-S or SWS spectra) divided by the 7.7µm continuum flux obtained from a linear extrapolation from both long (~11.1µm) and short (5.9µm) wavelengths. Data are from this paper, Moorwood et al.(1996: Circinus), Lutz et al.(1997:NGC 1068), Kunze et al.(1997: M82), Verstraete et al. (1996: M17), Vigroux et al. (1996, NGC 4038/39). Starburst templates (plus two positions in the Galactic center and in M17) are marked as open triangles, ULIRGs as filled circles and AGNs as crossed rectangles. Circinus, Cen A and NGC 7582 are composite sources with a central AGN dominating but star formation contributing and that NGC 7469 is a starburst galaxy with a Seyfert 1 nucleus contributing. A typical error bar (for NGC 1068) is plotted.

Fig.5. A new diagnostic diagram combining the extinction corrected 26µm [OIV]/12.8µm [NeII] (or alternatively 1.7x(33.4µm [SIII])) line ratio (vertical) with the strength of the 7.7µm PAH feature (horizontal). Starburst galaxies are marked as open triangles, ULIRGs as filled circles and AGNs as crossed rectangles. Downward arrows denote upper limits (derived from 'most probable' errors) and a typical error (for NGC 1068) is denoted by a cross. Dashed arrows at $45^o$ denote where composite sources would move if their observed charcteristics were corrected for the starburst component (Cen A, Circinus and N 7582), or their AGN component (N 7469). Left: basic data with individual sources marked. Right: shown  dashed (with crosses) is a simple linear 'mixing' curve of combining various fractions of total luminosity in an AGN (numbers marked) and in a starburst. 100% AGN is assumed to be[OIV]/[NeII]~1, PAH strength~0.04, 0% AGN (=100% starburst) is assumed to be [OIV]/[NeII]~0.02, PAH strength~3.6. The areas of the diagram dominated by starformation and by AGNs are denoted.

Fig.6. Histograms of number of sources as a function of the ratio of 8-1000μm infrared luminosity to Lyman continuum luminosity ($L_{IR}/L_{Lyc}$). Top: ULIRGs from Goldader et al. (1995), plus UGC 5101 and NGC 6240, with L(Lyc) from Brγ and near-infrared/optical extinction corrections. Middle: Starburst galaxies from our ISO data (Table 2). Bottom: ULIRGs from ISO data with extinction corrections as in Table 2.

Fig.7. Histograms of number of sources as a function of the ratio of K-band luminosity to Lyman continuum luminosity ($L_K/L_{Lyc}$, $L_{Lyc}$ from Brγ). Top: Starbursts. Middle: ULIRGs. Bottom: AGNs. In all cases the sources are taken from Goldader et al. (1995, 1997a,b), Moorwood and Oliva (1988)and Genzel et al.(1995).

Fig.8. 3D maps of UGC 5101 in the K-band (0.9" FWHM resolution). Top left: K-band (blue end: ~2.1μm) with contours 0.025, 0.05, 0.075, 0.1, 0.125, 0.15, 0.2, 0.3 .... 0.9, 0.95 of peak flux. Top middle: Pα flux (contours 0.15, 0.25, 0.35 ... 0.95 of peak flux) superposed on greyscale of K-band (blue end). Top right: sum of $H_2$ S(1) and S(3) (contours 0.15, 0.25 0.35 ... 0.95 of peak) superposed on grey scale of K-band. Bottom left: K-band (red end: ~2.35μm) with contours 0.025, 0.05, 0.075, 0.1, 0.125, 0.15, 0.2, 0.3 .... 0.9, 0.95 of peak flux. Bottom middle: Paα channel maps +300 km/s and -300 km/s in contours (0.25, 0.5, 0.75, 0.95 of peak) and 0 km/s in greyscale. Bottom right: CO 0-2 absorption flux (contours 0.25, 0.45 0.65, 0.85, 0.95 of peak flux) superposed on greyscale K-band (red end).

Fig.9. CO H-band absorption flux (grey scale) superposed on H-band contours (0.025, 0.05, 0.075, 0.1, 0.125, 0.15, 0.2, 0.3 .... 0.9, 0.95 of peak flux, 0.7" FWHM). The asterisk marks the position of the continuum peak.

Fig.10. +300/-300 km/s Pα channel maps (contours) superposed on a grey-scale version (in logarithmic units) of the 0.2" resolution, 8.4 GHz radio continuum map of Condon et al. (1991b). For alignment the radio continuum peak is assumed to be coincident with the K-band continuum peak.

Fig.11. H-band (left) and K-band (right) 3D spectra of different positions in UGC 5101. The positions from top to bottom are 1.5" east, nucleus and 1.5" west of the nucleus, each with an aperture of about 1-1.5".

Fig.12. Top(A): Sample $^{12}$CO J=1-0 IRAM Plateau de Bure interferometer spectra from the central 6" of UGC 5101, on a 1" grid. The spectral resolution is 20 km/s. The position offset from the interferometer phase center (near the nucleus) in arcsec is denoted in the upper right of each box. Bottom left (B): CO channel maps from the channels at relative velocities of +240 km/s (solid contours) and -240 km/s (dashed contours). The lowest contour is at 6 mJy/beam ($3\sigma$) with levels increasing in steps of 6mJy/beam. The + indicates the interferometer phase center. Bottom left ( C): Contour map of integrated CO flux. The contours are logarithmically spaced by 0.15 in the $\log_{10}$, with the lowest level at 0.5 Jy km/s. The 2.1"x1.6" FWHM synthesized beam is shown in the lower left corner, and the + indicates the interferometer phase center. Bottom right (D): Observed position-velocity CO map taken along the kinematic major axis (p.a. 79$^\circ$). Contour levels start at 6 mJy/beam ($3\sigma$) and increase in steps of 6 mJy/beam.

Fig.13: Global starburst parameters. Shown are $L_{bol}/L_{Lyc}$, $L(K)/L_{Lyc}$, $10^{12}$ $\nu_{SNR}/L_{bol}$ and M(stars)/L(K) as a function of time for Salpeter IMFs with an upper mass cutoff of 50 $M_o$, lower

mass cutoff of 1 $M_\odot$ and for burst durations $\Delta t=10^6$ years (long-short dash), $2\times10^7$ years (triple dash) and $10^9$ years (long dash). The mean values of starburst galaxies and ULIRGs are denoted by fat dashed and continuous horizontal lines, respectively. The thin horizontal lines in the $L_{bol}/L_{Lyc}$ curve denote the dispersion in the ULIRG values. The location of UGC 5101 in all plots is indicated. Its range of possible ages marked as vertical lines in the M(star)/L(K) plot.